\documentstyle[12pt,epsf,righttag]{amsart}
\newtheorem{thm}{Theorem}[section]
\newtheorem{prop}[thm]{Proposition}

\newtheorem{lem}[thm]{Lemma}

\newtheorem{remark}[thm]{Remark}

\newcommand{\bbR}{{\Bbb{R}}}

\newcommand{\calB}{{\cal{B}}}
\newcommand{\calC}{{\cal{C}}}
\newcommand{\calD}{{\cal{D}}}
\newcommand{\calE}{{\cal{E}}}
\newcommand{\calF}{{\cal{F}}}
\newcommand{\calH}{{\cal{H}}}
\newcommand{\calI}{{\cal{I}}}

\newcommand{\calM}{{\cal{M}}}
\newcommand{\calN}{{\cal{N}}}
\newcommand{\calP}{{\cal{P}}}

\newcommand{\calS}{{\cal{S}}}

\newcommand{\calZ}{{\cal{Z}}}

\newcommand{\bzero}{{\bold 0}}

\newcommand{\bX}{\bold{X}}
\newcommand{\br}{\bold{r}}
\newcommand{\bR}{\bold{R}}
\newcommand{\wu}{\widehat{u}}
\newcommand{\ws}{\widehat{\sigma}}
\newcommand{\wf}{\widehat{f}}
\newcommand{\wa}{\widehat{\alpha}}
\newcommand{\wb}{\widehat{\beta}}
\renewcommand{\wp}{\widehat{p}}
\newcommand{\wq}{\widehat{q}}
\renewcommand{\wr}{\widehat{r}}
\newcommand{\wt}{\widehat{t}}
\newcommand{\tf}{\tilde{f}}
\newcommand{\ts}{\tilde{s}}
\newcommand{\tcs}{\tilde{\cal S}}

\newcommand{\cm}{\operatorname{cm}}
\newcommand{\lb}{\label}
\begin{document}
\title[Gravitational Ionization]{Gravitational Ionization: \\
Periodic Orbits of Binary Systems\\
Perturbed by Gravitational Radiation}
\author{C.~Chicone}
\thanks{The first author's research was supported by
the National Science Foundation under the grant  DMS-9303767}
\author{B.~Mashhoon}
\author{D. G.~Retzloff}
\date{\today}
\maketitle
\begin{abstract}  The long term perturbation of a Newtonian binary
system by an incident gravitational wave is discussed in connection
with the issue of gravitational ionization.
The periodic orbits of the planar tidal
equation are investigated and the
conditions for their existence are presented.
The possibility of ionization of a Keplerian orbit via gravitational
radiation is discussed.
\end{abstract}
\section{\lb{intro}Introduction}
A Newtonian two-body system cannot be completely
isolated from all other masses
in the universe as a consequence of the universality of the gravitational
interaction.  In fact, the attraction of the other masses would cause the
binary system to move through approximately inertial spacetime.
This center-of-mass motion should be distinguished from the
relative motion, which is affected by the gradient of the disturbing
forces.  Consider, for instance,
the equations of motion for an ``isolated'' two-body system in Newtonian
mechanics
\begin{align}
m_1 \frac{d^2 {\bX}_1}{dt^2} + \frac{G_0 m_1m_2}{|{\bX}_1 - {\bX}_2|^3}
({\bX}_1 - {\bX}_2)  = & - m_1 \nabla \Phi({\bX}_1),\notag \\
m_2 \frac{d^2 {\bX}_2}{dt^2} + \frac{G_0 m_1m_2}{|{\bX}_1 - {\bX}_2|^3}
({\bX}_2 - {\bX}_1)  = & - m_2 \nabla \Phi({\bX}_2),\lb{BasicEq1}
\end{align}
where $G_0$ is Newton's constant of gravitation and
\[\Phi({\bX}):=-\sum_p\frac{G_0m_p}{|{\bX}-{\bX}_p|}\]
represents the combined gravitational potential of all
other masses $m_p$ at ${\bX}_p$ in the universe. Here and throughout this work,
the finite size of astronomical bodies is neglected.
If, in the inertial space coordinates $(X^1,X^2,X^3)$,
the binary system is so far away from the other masses that the
relative distance between the masses comprising the binary
is very small compared to the distance of
the center of mass of the binary to the external masses,
then, to first order in this small ratio, the equation of relative motion
has the form
\begin{equation}\lb{beqn}
\frac{d^2  r^i}{dt^2} + \frac{k  r^i}{r^3} = - K_{ij}(t) r^j,
\end{equation}
where ${\br}=(r^1,r^2,r^3): = {\bX}_1 - {\bX}_2$,
$r$ is the length of ${\br}$, and
$k$ = $G_0(m_1 + m_2)$. Here,
$K_{ij}$, the tidal matrix, is given by
\[
K_{ij} = \frac{\partial^2 \Phi}{\partial X^i \partial X^j}
\]
evaluated at the center of mass of the binary system.
In Newtonian mechanics the gravitational potential
$\Phi$ is a harmonic function; therefore, the symmetric
tidal matrix is trace-free.

It turns out that \eqref{beqn} holds approximately in general
relativity as well, except that $K_{ij}$ would be represented by the
``electric" components of
the Riemannian curvature of the underlying spacetime projected
onto a Fermi frame along the center-of-mass worldline
\cite{mashoon1,mashoon2}.  That is,
the equation of relative motion can be considered to be the
Newton-Jacobi equation in the sense that once the
internal Newtonian attraction is
neglected, equation~\eqref{beqn} reduces to the Jacobi equation in Fermi
normal coordinates for the
relative motion of two neighboring geodesics in the underlying
spacetime manifold.
Thus the spacetime coordinates in \eqref{beqn} refer to a local
Fermi system established along
the path of the center of mass of the system.
In our approximate treatment,
we neglect relativistic effects in the binary system.
On the other hand, the
external influences may now include gravitational radiation.
It should be noted in this connection that classical celestial mechanics has
been mainly concerned with the $n$-body problem; however, the
``vacuum'' between these bodies is expected to abound with gravitational
radiation as well as with other radiation fields. It is therefore interesting
to consider the interaction of gravitational waves with $n$-body systems, since
it is estimated that half of all stars are members of binary or multiple
systems.

In this paper, attention is focused on a Newtonian binary system that
undergoes perturbation due to an incident gravitational wave.  Let the
spacetime metric due to the gravitational wave be given by
\[
g_{\mu\nu} = \eta_{\mu\nu} + \epsilon\chi_{\mu\nu},
\]
where $\eta_{\mu\nu}$ is the Minkowski
metric, $\epsilon$ is the strength of the perturbation, $0<\epsilon\ll 1$,
and $\chi_{\mu\nu}$ represents the gravitational radiation field.
In the transverse traceless gauge, $\chi_{0\mu}$ = 0 and $\chi_{ij}$
is a symmetric traceless matrix that satisfies the wave equation
$\Box^2 \chi_{ij} = 0$ and the transversality condition
$\partial_j \chi_{ij} = 0$.
It turns out that in this gauge,
\begin{equation}\label{bbeqn}
K_{ij}(t) =
- \frac{1}{2}\epsilon \frac{\partial^2 \chi_{ij}}{\partial t^2}
    (t,{\bX}_{\cm}),
\end{equation}
where $(m_1 + m_2){\bX}_{\cm} = m_1{\bX}_1 + m_2{\bX}_2$.
It is possible to fix the position of the center of mass (e.g.,
$\bX_{\cm}=\bzero$) in the approximation under consideration here,
since $K_{ij}$ is considered only to first order in $\epsilon$. The
perturbing field $\chi_{ij}$ may be expressed as a Fourier sum
of plane monochromatic waves with wavelengths much larger than
the semimajor axis of the binary system. Such waves could be generated
by the motion of masses during the Hubble expansion, or could be
primordial waves left over from the big bang era. It is therefore
important to note that in our analysis there is no need to specify
the {\em initial conditions} for the interaction of the
waves with the binary; instead, we concentrate on the
``steady-state'' situation involving a dominant plane wave
of frequency $\Omega$ incident on the binary. Hence, the symmetric
and traceless tidal matrix in equation~\eqref{beqn} is given
in Cartesian coordinates by
\begin{align*}
{K}_{11}  = & \epsilon\alpha \Omega^2 {\cos^2{\Theta}} \cos{(\Omega t)}, \\
{K}_{12}  = & \epsilon\beta \Omega^2 \cos{\Theta} \cos{(\Omega t + \rho)}, \\
{K}_{13}  = &-\epsilon \alpha \Omega^2 \cos{\Theta}
                       \sin{\Theta} \cos{(\Omega t)}, \\
{K}_{22}  = & -\epsilon\alpha \Omega^2 \cos{(\Omega t)}, \\
{K}_{23}  = & -\epsilon\beta \Omega^2\sin{\Theta} \cos{(\Omega t + \rho)},
\end{align*}
where $\alpha$, $\beta$, and $\rho$ are constants,
and $\Theta$ is the polar angle from the normal to the plane of the
unperturbed orbit to the propagation vector of the incident radiation.
However, for the sake of simplicity we will
explicitly consider only the case of normal incidence, i.e. $\Theta$ = 0.
In this case, the orbital plane will remain fixed; that is,
when the waves are normally incident, the problem reduces to planar
motion under the approximations considered here.

Let the unperturbed Keplerian motion be confined to the $(x,y)$-plane.
The transverse nature of the radiation
field implies that the orbital plane will be unchanged
under the perturbation of normally incident waves.
Thus in \eqref{beqn},
we have $r^1 = x$, $r^2 = y$, and $r^3 = 0$.
The nonzero elements of the tidal matrix for our single-frequency
radiation are
\begin{alignat*}{2}
K_{11} & =&  - K_{22} =& \epsilon\alpha \Omega^2 \cos{(\Omega t)},\\
K_{12}& = &K_{21} =& \epsilon\beta \Omega^2 \cos{(\Omega t +  \rho)},
\end{alignat*}
where $\epsilon\alpha$ and $\epsilon\beta$  represent the
amplitudes of the two independent linear polarization states of
the low-frequency gravitational wave,
and $\rho$ represents the constant phase difference between them.

The justification for replacing the actual problem with this
rather simplified nonlinear model is that it becomes amenable to
mathematical analysis. It should also be remarked that---within the
limitations discussed in this section---equation~\eqref{beqn}
for the relative motion holds
generally in the Fermi coordinate system established along the
center-of-mass worldline.  Thus, for this system,
$K_{ij} = R_{0i0j}$, where $R_{\mu\nu\rho\sigma}$ denotes the Riemannian
curvature due to external sources projected onto the frame
of the center of mass.
In the Newtonian limit of general relativity, each $K_{ij}$
reduces to a second partial derivative of the external
Newtonian potential $\Phi({\bX})$ evaluated along the
path of the center of mass.
On the other hand, for a weak gravitational wave,
equation~\eqref{beqn} holds to first order in the amplitude of the
gravitational potential.
Thus, in general,
the matrix $(K_{ij})$ is a function of the proper time along the path of
the center of mass.
It is always possible to diagonalize this symmetric matrix;
however, its dependence upon time implies that \eqref{beqn}
must then be written in a rotating system of coordinates.

In electrodynamics, the interaction of electromagnetic radiation with a
two-body system constitutes a basic problem (e.g.,  the scattering and
absorption of light by the Rutherford-Bohr atom).
The gravitational analog of
this problem in the classical regime would involve the scattering and
absorption of gravitational radiation by a Keplerian two-body system.
The wavelength of light is much larger than the Bohr radius of the atom;
therefore, the dominant interaction takes place via the
electric dipole moment
of the atom since electromagnetism is a spin - 1  field.
We expect by analogy
that for gravitational radiation with a (reduced) wavelength
that is much larger
than the semimajor axis of the Keplerian orbit, the dominant
interaction would involve the mass quadrupole moment of the binary since
gravitation is a spin - 2 field.
This approximation corresponds precisely to
dropping higher-order terms in the tidal equation~\eqref{beqn}.

The reciprocity between emission and absorption of
radiation should be noted.
In the quadrupole approximation for the emission of
gravitational radiation, the
waves carry away energy and angular momentum but not linear momentum.
The same holds in the inverse process as well,
except that in general the system can gain or lose energy.
Moreover, a Newtonian binary system emits gravitational radiation
of frequency $\Omega=m\omega$, $m=1,2,3\ldots$, where $\omega$
is the Keplerian frequency of the elliptical orbit. Similarly,
resonant absorption of gravitational waves by an elliptical
binary occurs at $\Omega=m\omega$, $m=1,2,3\ldots$, according
to the linear perturbation analysis~\cite{mashoon2}.

A two-body system continuously emits gravitational radiation according to
general relativity.  Gravitational energy in the form of radiation is thus
carried away from the system.  Hence, the relative orbit evolves in such a way
that the semimajor axis of the osculating ellipse monotonically shrinks.
This phenomenon of inward spiraling of the members of the binary is
consistent with the timing observations of
the Hulse-Taylor binary pulsar~\cite{ht}~\cite{ht1}.
Direct observational evidence for gravitational radiation does not
exist at present; however, efforts are under way to detect in
the laboratory gravitational waves emitted by astrophysical sources with
$\epsilon\approx 10^{-20}$.

In this work, we will ignore the emission of gravitational
radiation by the binary
system, and  concentrate our attention instead on the absorption
process.  The flow of energy between the incident radiation and the binary is
not unidirectional, however.  The self-gravitating system can absorb energy
from the radiation field or deposit energy into the wave so as to induce an
amplification of the radiation. These issues were first discussed
in connection with the problem of ionization~\cite{mashoon2}
in the context of linear perturbation analysis that in general
breaks down over time.  Here we employ the concepts introduced
by Poincar\'e~\cite{hp} for the treatment of nonlinear problems.
These enable us to
prove that periodic orbits exist in the perturbed system for which
energy must steadily flow back and forth between the wave and the binary.
It is
important to emphasize that such periodic orbits occur near resonance
conditions when certain definite phase relationships are satisfied.
If the binary system monotonically absorbs energy from the wave, then the
semimajor axis of the osculating ellipse will grow with time and the system
eventually ionizes.  We provide a qualitative picture for such a process in
\S~\ref{scn}.

The gravitational quadrupole interaction may be
illustrated by considering the
Hamiltonian for the relative motion
\begin{equation}\label{bhamiltonian}
{\cal H} = \frac{1}{2} p^2 - \frac{k}{r} + \frac{1}{6} K_{ij}(t) Q_{ij},
\end{equation}
where the quadrupole moment per unit mass is defined by
$Q_{ij} = 3 r^i r^j - r^2 \delta_{ij}$ and in the most general case
of a normally incident gravitational wave packet considered
in this paper the matrix $K_{ij}$
is a traceless symmetric matrix of periodic functions.
Thus  $K_{11} = - K_{22} = h_1(t)$,
$K_{12} = K_{21} = h_2(t)$, and there exist $\tau_1$ and $\tau_2$
such that $h_1(t) = h_1(t + \tau_1)$ and $h_2(t) = h_2(t + \tau_2)$.
Here $h_1$ and $h_2$ represent the amplitudes of the two
linearly independent polarization
states of the perpendicularly incident gravitational wave.
Now let
\begin{equation*}
{\cal E}  =  \frac{1}{2} p^2 - \frac{k}{r}, \quad
{\cal L}^i  =  \epsilon_{ijk} r^j p^k, \quad
\eta^i  =  \epsilon_{ijk} p^j {\cal L}^k - \frac{k}{r} r^i
\end{equation*}
denote the Newtonian energy, orbital angular momentum and the
Runge-Lenz vector associated with the relative motion of the
system (per unit reduced mass); then, these otherwise conserved
quantities vary as a consequence
of the coupling of the quadrupole moment of the system to the
curvature of the background spacetime.
Thus, in this quadrupole approximation, the Keplerian
orbit exchanges energy and angular momentum with the radiation field.
At each instant,  the relative motion can be described by the
orbital elements of the
osculating ellipse. This osculating ellipse continuously makes transitions
to other osculating Keplerian orbits with different energies and
angular momenta as a consequence of interaction with the external wave.
It is interesting to describe the path of
the system in the six-dimensional manifold of orbital elements;
in fact, this paper is
devoted to the description of periodic paths in this manifold.

The interaction of gravitational radiation with
matter may have played a
significant role in the evolution of the universe.
The treatment presented here
is confined, however, to the interaction of an incident wave
with a {\em Newtonian} binary system. In particular, we neglect all
relativistic effects
in the relative motion of the binary system.
Let the system have a Keplerian frequency $\omega$ and semimajor
axis $a$; then, by Kepler's third law, $\omega^2=k/a^3$.
Relativistic two-body effects may be neglected
provided $k\ll c^2 a$, where $c$ is the speed of light in vacuum.
Moreover, the quadrupole approximation for the interaction of the system
with the gravitational wave is valid if $\Omega a\ll c$.
More generally, our approach is sound provided
\[
\big(\frac{k}{c^2 a}\big)^{1/2}\ll\frac{\Omega}{\omega}
\ll\big(\frac{k}{c^2a}\big)^{-1/2}.
\]
Furthermore, the requirement that the external wave be a small
perturbation of the system implies that
$\Omega/\omega\ll 1/\sqrt{\epsilon }$,
since the strength of the interaction is given by
$\epsilon\Omega^2/\omega^2$, a quantity that must be much smaller than
unity.
Our objective is to determine the
conditions under which periodic Keplerian motions of the binary are
continued to periodic motions under the interaction.
It is an important consequence of our methods, which are originally due
to Poincar\'e~\cite{hp}, that the existence of higher-order perturbing
influences on the orbit, i.e., terms of order at least $\epsilon^2$,
can only affect the shape of the resulting periodic orbit but not its
existence.
In this first treatment of the nonlinear case,
we consider only special cases of the other
interesting questions---such as gravitational
ionization---that are suggested by the electrodynamic
analogy discussed in \cite{mashoon2}.
A treatment of the general problem on the basis of
linear perturbation theory is
contained in previous work \cite{mashoon2}.
Superposition of linear
perturbations due to the Fourier components of a
pulse of gravitational
radiation permits a general analysis in that case;
however, the validity of the
treatment is restricted in time as temporal evolution leads
to a breakdown of
the linear perturbation theory.
Therefore, the intrinsic nonlinearity of the
system must in general be taken into account for applications
in celestial mechanics. In this regard, we mention that the equations
of motion of a binary influenced by the gravitational attraction
of a massive distant  third body
can also be treated using the methods developed here; this constitutes
a special limiting case of the three-body problem and is discussed
in Appendix B.

We will be mainly concerned with
the continuation of Keplerian orbits under perturbation
by a resonant gravitational wave. In general, we show that
all but  a  finite
set of such resonant orbits are not  continued to periodic orbits under
the influence of the incident wave and that in general all elements of
the finite exceptional set are, in fact, continued.

The plan of this paper is the following. In \S~\ref{de}, we transform
the perturbation problem to Delaunay elements and obtain explicit
expressions for the transformed perturbation in terms of Fourier
series with coefficients that involve the Bessel functions.
In \S~\ref{ct}, we outline a continuation
method based on the Lyapunov-Schmidt reduction that is
adapted from \cite{ccc}.
In \S~\ref{cko}, the results of the continuation theory are applied
to the perturbation
of a binary influenced by a normally
incident wave. In particular, we find  bifurcation equations
for the problem, that is, a system of equations whose simple zeros correspond
to the continuable periodic Keplerian orbits and we show that
these equations indeed have simple zeros.
In \S~\ref{cpw}, we consider the special case of circularly polarized
gravitational waves, a  case that is not covered by the
results of \S~\ref{cko}. In this case we show that there are periodic
solutions.
We also show for sufficiently weak perturbations of Keplerian ellipses
that the semimajor axis of the osculating ellipse
remains bounded for all time.
The final section, \S~\ref{scn}, contains a brief discussion
of some additional speculative results on the ionization problem.
Some standard formulas are relegated to Appendix~A, and
in Appendix~B we consider a special case of the three body problem:
A binary influenced by a distant massive third body.
\section{\lb{de}Delaunay Elements and Fourier Series Expansion}
In terms of the canonical variables $(p_r,p_\theta,r,\theta)$, that is
\[
x=r\cos\theta,\quad y=r\sin\theta,\quad p_r=\dot r,\quad p_\theta=
r^2\dot\theta,
\]
the Hamiltonian for our perturbation problem, with perturbation
parameter $\epsilon$,  may be expressed in the
following general form
\begin{equation}\lb{perHam}
\calH(p_r,p_\theta,r,\theta)=
    \frac{1}{2}(p_r^2+\frac{p_\theta^2}{r^2})-\frac{k}{r}
+\epsilon r^2 (\phi(t)\cos{2\theta}+\psi(t)\sin{2\theta}),
\end{equation}
where $\phi$ and $\psi$ are periodic functions with a common period.
We will continue to
use this general form in order to show how our theory can be applied.
However, for the sinusoidal monochromatic gravitational wave
model we will consider only the case where
\begin{equation}\lb{phipsi}
\phi(t)=\frac{1}{2}\alpha\Omega^2\cos(\Omega t),\qquad
\psi(t)=\frac{1}{2}\beta\Omega^2\cos(\Omega t+\rho).
\end{equation}
The unperturbed Hamiltonian
\begin{equation}\lb{KepHam}
H(p_r,p_\theta,r,\theta)=\frac{1}{2}(p_r^2+\frac{p_\theta^2}{r^2})-\frac{k}{r}
\end{equation}
is called the Kepler Hamiltonian. We will consider only those
motions corresponding to negative energy $E=H(p_r,p_\theta,r,\theta)$.

Following S.~Sternberg~\cite[Vol. 2, pp. 234-247]{Stern},
we define
\[ L:=\big(\frac{-k^2}{2E}\big)^{1/2},\qquad G:=p_\theta,\]
and we let $a(1\pm e)$ denote the roots of the quadratic
polynomial
\[
r^2-\frac{2L^2}{k}r+\frac{G^2L^2}{k^2}=0
\]
so that
\begin{equation}\lb{ea}
a=\frac{L^2}{k},\qquad e=\frac{1}{L}(L^2-G^2)^{1/2}.
\end{equation}
Here,
$a$ is the semimajor axis and $e$ is the eccentricity of the Keplerian ellipse
with $0\le e<1$.
However, we will only consider the case
$e>0$, that is noncircular orbits.
With this restriction in force,
we define  $\wu$, the eccentric anomaly and $v$, the true anomaly,
implicitly by the formulas
\begin{equation}\lb{anomaly}
r=a(1-e\cos \wu),\qquad r=\frac{a(1-e^2)}{1+e\cos v},
\end{equation}
and new variables $\ell$ and $g$ by
\[
\ell=\wu-e\sin \wu,\qquad g=\theta-v.
\]

As proved in \cite{Stern}, the change of coordinates
\[
(p_r,p_\theta,r,\theta)\to (L,G,\ell,g)
\]
is canonical. Here $\ell$ and $g$ are ``angle variables'', defined
modulo $2\pi$, while $L$ and $G$ are ``action variables''. The
new coordinates $(L,G,\ell,g)$ are called the Delaunay elements.

In Delaunay variables, our Hamiltonian~\eqref{perHam}
is transformed to
\[
\calH=
-\frac{k^2}{2L^2}+\epsilon(\calC(L,G,\ell,g)\phi(t)+\calS(L,G,\ell,g)\psi(t)),
\]
where $\calC$ (resp.\ $\calS$) is the function obtained by
expressing $r^2\cos{2\theta}$
(resp. $r^2\sin{2\theta}$) in terms of the
Delaunay elements. Using the fact that the change of coordinates
is canonical, the differential equations of motion are given
by the Hamiltonian system
\begin{alignat}{2}\lb{eqmotion}
\notag \dot L=&&-&\epsilon(
               \frac{\partial \calC}{\partial\ell}(L,G,\ell,g)\phi(t)+
                \frac{\partial \calS}{\partial\ell}(L,G,\ell,g)\psi(t)),\\
\notag \dot G=&&-&\epsilon(
               \frac{\partial \calC}{\partial g}(L,G,\ell,g)\phi(t)+
                \frac{\partial \calS}{\partial g}(L,G,\ell,g)\psi(t)),\\
\notag \dot \ell=&&\omega+&\epsilon(
               \frac{\partial \calC}{\partial L}(L,G,\ell,g)\phi(t)+
                \frac{\partial \calS}{\partial L}(L,G,\ell,g)\psi(t)),\\
\dot g=&&&\epsilon(
               \frac{\partial \calC}{\partial G}(L,G,\ell,g)\phi(t)+
                \frac{\partial \calS}{\partial G}(L,G,\ell,g)\psi(t)),
\end{alignat}
where $\omega:=k^2/L^3$ is the frequency of the elliptical Keplerian
orbit.

In order to analyze system~\eqref{eqmotion}, we must find computable
expressions for the partial derivatives of $\calC$ and $\calS$.
This can be done in several
ways; however, for our purposes, the most useful expressions are obtained
from Fourier series expanded as functions of the angle
variable $\ell$.
The determination of these series is outlined in Appendix A, and the result
can be expressed as
\begin{align}
\notag\calC(L,G,\ell,g)=&\frac{5}{2}a^2e^2\cos{2g}
  +a^2\sum_{\nu =1}^\infty (A_\nu\cos{2g}\cos {\nu\ell}
                            -B_\nu\sin{2g}\sin{\nu \ell}),\\
\lb{CSfs}\calS(L,G,\ell,g)=&\frac{5}{2}a^2e^2\sin{2g}
  +a^2\sum_{\nu =1}^\infty (A_\nu\sin{2g}\cos {\nu\ell}
                            +B_\nu\cos{2g}\sin{\nu \ell}),
\end{align}
where
\begin{align}\lb{fcoeff}
\notag A_\nu=&\frac{4}{\nu^2e^2}
            (2\nu e(1-e^2)J_\nu'(\nu e)-(2-e^2)J_\nu(\nu e)),\\
B_\nu=&-\frac{8}{\nu^2e^2}\sqrt{1-e^2}\,
     (e J_\nu'(\nu e)-\nu (1-e^2)J_\nu(\nu e)).
\end{align}

\section{\lb{ct}Continuation Theory}
In order to analyze the continuation (persistence) of periodic solutions
of the Kepler system to system~\eqref{eqmotion}, we use a method
proposed in~\cite{ccc}. Here, we outline the main ideas; the reader
is referred to~\cite{ccc} for the details.

Consider a system of the  form
\begin{equation}
\lb{bifeq}
\dot u=F(u)+\epsilon h(u,t),
\end{equation}
where $u$ is a coordinate on a manifold $M$ consisting of a
cross product of Euclidean spaces and
tori, $h$ is $2\pi/\Omega$ periodic in its second variable, and
$\epsilon$ is a small parameter.
Let $t\mapsto u(t,\xi,\epsilon)$ denote the solution
of~\eqref{bifeq} with initial condition $u(0,\xi,\epsilon)=\xi$,
$\xi\in M$. Also, we define the $m$th order Poincar\'e map by
$\calP^m(\xi,\epsilon)=u(2m\pi/\Omega,\xi,\epsilon)$;
it corresponds to a strobe that illuminates the orbit after $m$ cycles
of the perturbation. Of course, a fixed point of
$\xi\mapsto \calP^m(\xi,\epsilon)$
corresponds to a periodic orbit of~\eqref{bifeq}.
If $m$ is the smallest such integer for which $\xi$ is a fixed point,
then $\xi$ is the initial point of a subharmonic of order $m$.

Suppose that there is a submanifold $\calZ\subset M$ consisting entirely
of fixed points of the unperturbed order $m$ Poincar\'e map, defined by
$p^m(\xi):=\calP^m(\xi,0)$.
Our continuation theory is a method, one among many, to decide if
any of these fixed points survive after perturbation. More precisely,
we say a point $z\in \calZ$, and therefore the unperturbed periodic
orbit of~\eqref{bifeq} with initial point $z$,
is continuable (or that it persists) if
there is a continuous curve $\epsilon\mapsto \gamma(\epsilon)$ in $M$
such that
$\gamma(0)=z$ and
$\calP^m(\gamma(\epsilon),\epsilon)\equiv \gamma(\epsilon)$.
Here, $\gamma(\epsilon)\in M$ is the initial point of a periodic
solution of~\eqref{bifeq}.

In order to apply the method of~\cite{ccc}, namely Lyapunov-Schmidt
reduction to the Implicit Function Theorem, the fixed-point manifold
(resonance manifold) $\calZ$ must satisfy a nondegeneracy condition
relative to the unperturbed Poincar\'e map. To specify this condition,
consider $z\in \calZ$ and  the derivative  $Dp^m(z)$ viewed as a linear
transformation of the tangent space $T_zM$. The base point stays fixed
because $p^m$ is the identity on $\calZ$. Moreover, every vector
in $T_zM$ that is tangent to the submanifold $\calZ$ is fixed by
$Dp^m(z)$, or, as we will say, every such vector is in the kernel
of the infinitesimal displacement $\calD(z)= Dp^m(z)-I$. The manifold
$\calZ$ is called normally nondegenerate if the kernel of the infinitesimal
displacement is exactly the tangent space $T_z\calZ\subset M$.
Equivalently, $\calZ$ is normally nondegenerate, if for each $z\in \calZ$,
the dimension of the kernel of the infinitesimal displacement at $z$
is equal to the dimension of the manifold $\calZ$.

Suppose $\calZ$ has dimension $\Delta$ and that it is a
normally nondegenerate submanifold of $M$. In this case the range of
the infinitesimal displacement at each point in $\calZ$ has codimension
$\Delta$.
Thus, for $z\in \calZ$, there is a vector space complement $\tcs(z)$,
to the range of $\calD(z)$. We let $\ts(z)$ denote the projection of
$T_zM$ to $\tcs(z)$. By choosing local coordinates, we note that
both $\calZ$ and $\tcs(z)$ may be identified with $\bbR^\Delta$.

Let $z\in \calZ$ and consider the curve  in $M$ given by
$\epsilon\mapsto \calP^m(z,\epsilon)$. This curve passes through $z$ at
$\epsilon=0$. Its tangent vector at $\epsilon=0$, which
may be identified with the partial derivative $\calP^m_\epsilon(z,0)$,
is in $T_zM$.
We define the bifurcation function $\calB$ to be the map,
from $\calZ$ to the complement $\tcs$ of the
range of the infinitesimal displacement, given by
\[
\calB(z)=\ts(z)\calP^m_\epsilon(z,0).
\]
In local coordinates $\calB:\bbR^\Delta\to\bbR^\Delta$.
We will say $z\in\calZ$ is a simple zero of the bifurcation function provided
$\calB(z)=0$ and the derivative $D\calB(z)$ is invertible.

A result in~\cite{ccc} is the following continuation theorem:
\begin{thm}\lb{conthm}
If $\calZ$ is a normally nondegenerate fixed-point
submanifold of $M$ for system~\eqref{bifeq} and if
$z\in \calZ$ is a simple zero of the associated bifurcation function,
then the unperturbed periodic orbit
of~\eqref{bifeq} with initial point $z$ is continuable.
\end{thm}

To use Theorem~\ref{conthm} as a practical tool, we must be able to
compute $\calP^m_\epsilon(z,0)$. Fortunately,
this partial derivative can usually be computed. In fact, if
$\Omega=\Omega(\epsilon)$, then
\[
\calP^m_\epsilon(z,0)=
-\frac{2m\pi}{\Omega(0)^2}\Omega'(0)\dot u(2m\pi/\Omega(0),z,0)
+u_\epsilon(2m\pi/\Omega(0),z,0).
\]
Thus,
if $t\mapsto W(t)$ is
the solution of the second variational initial value problem
\[
\dot W=DF(u(t,z,0))W+h(u(t,z,0),t),\quad W(0)=0,
\]
then
\[\calP^m_\epsilon(z,0)=
-\frac{2m\pi}{\Omega(0)^2}\Omega'(0)F(u(2m\pi/\Omega(0),z,0))+
W(2m\pi/\Omega(0)).\]
In effect,
$W(t)=u_\epsilon(t,z,0)$ with
$W(0)=0$ because $u(0,z,\epsilon)=z$.

In our gravitational
radiation model, it seems appropriate that the frequency
of the gravitational wave is independent of the amplitude
of the wave. Thus, we will assume below that $\Omega$ does not
depend on $\epsilon$. This simplifies the expression for
$\calP^m_\epsilon(z,0)$ by removing the
 ``detuning'' .
\section{\lb{cko}Continuation of Kepler Orbits}
To apply Theorem~\ref{conthm} to our perturbation problem~\eqref{eqmotion},
we must define a normally nondegenerate fixed-point manifold. For this,
we consider the Kepler orbits that are
in resonance with the periodic perturbation.

If there are fixed relatively prime positive integers $m$ and $n$
and a fixed value of $\omega$, the frequency of the Keplerian
orbit,  such that
\[
m\frac{2\pi}{\Omega}=n\frac{2\pi }{\omega},
\]
then the unperturbed solution of~\eqref{eqmotion} starting at
$(L,G,\ell,g)$ is given by
\[t\mapsto (L,G,\omega \wt+\ell,g),\]
where $\wt=t-t_0$ and $t_0$ is an integration constant that
denotes the starting instant of time.
A detailed analysis shows that $t_0$ can be set equal to zero here
without loss of generality.
Since $\ell$ is defined modulo $2\pi$, this solution is periodic
of period $2\pi /\omega$. Moreover, the $m$th order unperturbed
Poincar\'e map is defined by
\[
p^m(L,G,\ell,g)=(L,G,2\pi m\frac{\omega}{\Omega}+\ell,g).
\]

If we define the three-dimensional manifold
\[
\calZ^L:=\{(L,G,\ell,g): m\omega=n\Omega\},
\]
and recall that $\ell$ and $g$ are defined modulo $2\pi$, then
it follows immediately that $\calZ^L$ is fixed by $p$.
To check that $\calZ^L$ is normally nondegenerate, we compute
\[\calD(L,G,\ell,g)=Dp^m(L,G,\ell,g)-I=
\left(
\begin{array}{cccc}
0          &  0  &  0  &  0  \\
0          &  0  &  0  &  0  \\
-6\pi n/L  &  0  &  0  &  0  \\
0          &  0  &  0  &  0
\end{array}
\right)
\]
and we note that the infinitesimal displacement
has a three-dimensional kernel that is spanned by the usual basis vectors
\[
\left(
\begin{array}{c}
0  \\
1 \\
0 \\
0
\end{array}
\right),\quad
\left(
\begin{array}{c}
0  \\
0 \\
1 \\
0
\end{array}
\right),\quad
\left(
\begin{array}{c}
0  \\
0 \\
0 \\
1
\end{array}
\right).
\]
Moreover, the range of the infinitesimal displacement is complemented
by the span of the vectors
\[
\left(
\begin{array}{c}
1  \\
0 \\
0 \\
0
\end{array}
\right),\quad
\left(
\begin{array}{c}
0  \\
1 \\
0 \\
0
\end{array}
\right),\quad
\left(
\begin{array}{c}
0  \\
0 \\
0 \\
1
\end{array}
\right).
\]

To compute the bifurcation function associated with~\eqref{eqmotion},
we must compute the partial derivative
$\calP^m_\epsilon(G,L,g,\ell,0)$  on the manifold
$\calZ^L$ and then project the result into the complement
of the range of the infinitesimal displacement.
To do this, we simply solve the variational initial value problem
\begin{alignat*}{2}
\dot L_\epsilon=&&
       &-\frac{\partial \calC}{\partial\ell}(L,G,\omega t+\ell,g)\phi(t)-
                \frac{\partial \calS}{\partial\ell}(L,G,\omega
t+\ell,g)\psi(t),\\
\dot G_\epsilon=&&
       &-\frac{\partial \calC}{\partial g}(L,G,\omega t+\ell,g)\phi(t)-
                \frac{\partial \calS}{\partial g}(L,G,\omega
t+\ell,g)\psi(t),\\
\dot \ell_\epsilon=&&
       -\frac{3k^2}{L^4}L_\epsilon&+
       \frac{\partial \calC}{\partial L}(L,G,\omega t+\ell,g)\phi(t)+
                \frac{\partial \calS}{\partial L}(L,G,\omega
t+\ell,g)\psi(t),\\
\dot g_\epsilon=&&
       &\quad\frac{\partial \calC}{\partial G}(L,G,\omega t+\ell,g)\phi(t)+
                \frac{\partial \calS}{\partial G}(L,G,\omega t+\ell,g)\psi(t)
\end{alignat*}
with zero initial values and then project the solution computed
at $t=m2\pi/\Omega$ into the complement of the range of the infinitesimal
displacement.
{}From this procedure, we obtain the following bifurcation function
\[
\calB(G,\ell,g)=
(B^L(G,\ell,g),B^G(G,\ell,g),B^{g}(G,\ell,g)),
\]
where
\begin{equation}\lb{biffunc}
B^L(G,\ell,g):=-\frac{\partial \calI}{\partial\ell},
\quad
B^G(G,\ell,g):=-\frac{\partial \calI}{\partial g},
\quad
B^g(G,\ell,g):=\frac{\partial \calI}{\partial G},
\end{equation}
and
\[
\calI:=\int_0^{m 2\pi/\Omega}\big[
          \calC(L,G,\omega t+\ell,g)\phi(t)
+\calS(L,G,\omega t+\ell,g)\psi(t)\big] \,dt.
\]
Using the resonance relation, we have
\[
\calI=\int_0^{m 2\pi/\Omega}\big[
          \calC(L,G,\frac{n}{m}\Omega t+\ell,g)\phi(t)
+\calS(L,G,\frac{n}{m}\Omega t+\ell,g)\psi(t)\big]\,dt
\]
and, after changing the variable to $\ws=\Omega t/m+\ell/n$,
we obtain
\begin{align*}
\calI=&\frac{m}{\Omega}\int_{\ell/n}^{2\pi+\ell/n}\big[
\calC(L,G,n\ws,g)\phi(m(\ws-\ell/n)/\Omega) \\
&\hspace*{.4in}+\calS(L,G,n\ws,g)\psi(m(\ws-\ell/n)/\Omega)\big]\,d\ws.
\end{align*}
Using the fact that the last integrand is periodic with period $2\pi$
as a function of $\ws$ and substituting
$\phi$ and $\psi$ given in~\eqref{phipsi},
we find
\begin{align*}
\calI=&\frac{m\Omega}{2}\int_{0}^{2\pi}\big[
\alpha\calC(L,G,n\ws,g)\cos(m\ws-\frac{m\ell}{n}) \\
&\hspace*{.4in}+\beta\calS(L,G,n\ws,g)\cos( m\ws-\frac{m\ell}{n}+\rho)\big]
\,d\ws.
\end{align*}

To compute $\calI$, we substitute the Fourier series~\eqref{CSfs}
for $\calC$ and $\calS$ into the last expression for $\calI$ and use
trigonometric relations together with the fact that
$m$ and $n$ are relatively prime, to conclude that
$\calI=0$ unless $n=1$. Of course, there may be continuable
orbits for $n>1$, but they are not detected by our first
order method.  In case $n=1$, that is for the $(m:n)=(m:1)$ resonance,
we find that
\begin{equation*}
\begin{split}
\calI=\frac{1}{2}\pi ma^2\Omega\big(
\alpha(&A_m\cos{m\ell}\cos{2g}-B_m\sin{m\ell}\sin{2g})\\
+&\beta (A_m\cos(m\ell-\rho)\sin{2g}
          +B_m\sin(m\ell-\rho)\cos{2g})\big).
\end{split}
\end{equation*}
This result can be rewritten as
\begin{align}
\notag \calI=&\frac{1}{4}\pi m a^2\Omega\big[
(A_m+B_m)(\alpha-\beta\sin\rho)\cos(2g+m\ell) \\
\notag &+(A_m-B_m)(\alpha+\beta\sin\rho)\cos(2g-m\ell)
+\beta(A_m+B_m)\cos\rho\sin(2g+m\ell) \\
&+\beta(A_m-B_m)\cos\rho\sin(2g-m\ell) \big].
\lb{preB}
\end{align}
It is possible to express equation~\eqref{preB} in a more
compact form, in the usual manner, by defining
\begin{align}
\notag\calE\cos\sigma:=
   \beta\cos\rho,\quad &\calE\sin\sigma:=\alpha+\beta\sin\rho,\\
\lb{EF}\calF\cos\tau:=
   \beta\cos\rho,\quad &\calF\sin\tau:=\alpha-\beta\sin\rho,
\end{align}
so that
\[
\calE=(\alpha^2+\beta^2+2\alpha\beta\sin\rho)^{1/2},\quad
\calF=(\alpha^2+\beta^2-2\alpha\beta\sin\rho)^{1/2},
\]
and
\[
\calI=\frac{1}{4}\pi m a^2\Omega\big[
(A_m+B_m)\calF\sin(2g+m\ell+\tau)
+(A_m-B_m)\calE\sin(2g-m\ell+\sigma)\big].
\]
The simple zeros of the bifurcation function are then the same as the
simple zeros of
\begin{align}
\notag \calF(A_m+B_m)\cos(2g+m\ell+\tau)=&0,\\
\notag \calE(A_m-B_m)\cos(2g-m\ell+\sigma)=&0,\\
\lb{B}
\Big(\frac{\partial A_m}{\partial G}+\frac{\partial B_m}{\partial G}\Big)\calF
    \sin(2g+m\ell+\tau)+
\Big(\frac{\partial A_m}{\partial G}-\frac{\partial B_m}{\partial G}\Big)\calE
    \sin(2g-m\ell+\sigma)=&0.
\end{align}

To obtain explicit formulas for the partial derivatives of the
functions $A_m$ and $B_m$ with respect to $G$, we assume that
$G>0$ so that
\[\frac{\partial e}{\partial G}=-\sqrt{1-e^2}\,\frac{1}{Le}.\]
In case $G<0$, the partial derivative has the opposite
sign and the subsequent analysis is similar. We use
\eqref{A2},\eqref{A3},\eqref{A4} (from Appendix A), and \eqref{fcoeff}
to obtain
\begin{align}
\notag\frac{\partial A_m}{\partial G}=&-\sqrt{1-e^2}\,\frac{4}{L m^2 e^4}
  \big((2m^2(1-e^2)^2+4)J_m(me)-me(6-e^2)J_m'(me)\big),\\
\lb{parAB}\frac{\partial B_m}{\partial G}=& -\frac{8}{Lm^2e^4}
   \big(-3m(1-e^2)J_m(me)+e((2-e^2)(1-m^2e^2)+m^2)J_m'(me)\big).
\end{align}

The simple zeros of the function $\calB$ given by~\eqref{biffunc}
correspond to the continuable periodic orbits. Equivalently,
the continuable periodic orbits correspond to the simple solutions of the
system of equations given by~\eqref{B}. In order to find the
simple zeros of~\eqref{B}, we will use
the following proposition:
\begin{prop}\lb{gbaseprop}
If $\calE\calF\ne 0$ (that is, $|\alpha|\ne |\beta|$, or
$|\alpha|=|\beta|$  but $|\sin\rho| <1$) and if the system of
equations~\eqref{B} has a solution,
then \begin{equation}\lb{gbase}
(A_m^2-B_m^2)\Big\{
\Big(\frac{\partial A_m}{\partial G}+\frac{\partial B_m}{\partial G}\Big)^2
\calF^2-
\Big(\frac{\partial A_m}{\partial G}-\frac{\partial B_m}{\partial G}\Big)^2
\calE^2\Big\}
\end{equation}
is zero.
\end{prop}
\begin{pf}
If~\eqref{B} has a solution and
equation~\eqref{gbase} does not vanish, then, since the first factor
of equation~\eqref{gbase}
is not zero, we have
\[\cos(2g+m\ell+\tau)=0,\qquad \cos(2g-m\ell+\sigma)=0.\]
This implies that
\[\sin(2g+m\ell+\tau)=\pm 1,\qquad \sin(2g-m\ell+\sigma)=\pm 1,\]
and, since the third equation of~\eqref{B} is zero, that
\[
\Big(\frac{\partial A_m}{\partial G}
               +\frac{\partial B_m}{\partial G}\Big)\calF\pm
\Big(\frac{\partial A_m}{\partial G}
 -\frac{\partial B_m}{\partial G}\Big)\calE=0,
\]
in contradiction to the fact that~\eqref{gbase} is not zero.
\end{pf}

Proposition~\ref{gbaseprop}
reduces the search for solutions to several cases.
Just note that as soon as one of the factors of~\eqref{gbase}
vanishes, the value of $e$ and hence the
values of
\[
A_m-B_m,\quad A_m+B_m,\quad \frac{\partial A_m}{\partial G},\quad
   \frac{\partial B_m}{\partial G}
\]
are fixed. Thus,  equations~\eqref{B} reduce to solvable
trigonometric equations. It is important to note that
Proposition~\ref{gbaseprop} does not cover the interesting case
of circular polarization, which is therefore deferred to~\S~\ref{cpw}.

To study the zeros of~\eqref{gbase}
we will use the following proposition.
\begin{prop}\lb{zeroprop}
For the $(1:1)$ resonance, the functions $A_1+B_1$ and $A_1-B_1$
appearing in equations~\eqref{B} and viewed as functions of $e$
are both negative on the interval $0<e<1$. The functions
$\partial A_1/\partial G$ and $\partial B_1/\partial G$, viewed
as functions of $e$, each
have exactly one  simple zero on the interval and their zeros
are distinct.
For the $(m:1)$ resonance with $m>1$,
the range of the function
$(\partial A_m/\partial G)/(\partial B_m/\partial G)$, viewed
as a function of $e$ on the interval $0<e<1$, contains the interval
$[-1,1]$.
\end{prop}
\begin{pf}
We will outline the proof of the proposition. Some of the computations
were checked using a computer algebra system.

Consider the case $m=1$.
We will use the following elementary lemma~\cite[Lemma 3.5]{cj}
to  show that the function $f$ defined by
\begin{align}
\notag f(e):=&\frac{e^2}{4}(A_1-B_1)\\
\lb{fe}     =&2e((1-e^2)+\sqrt{1-e^2})J'_1(e)
               -((2-e^2)+2(1-e^2)^{3/2})J_1(e)
\end{align}
is negative on the interval $I_0:=\{e:0<e<1\}$.
\begin{lem}\lb{cjlem}
Suppose $f$ is a smooth function defined on an interval $[a,b)$
with the additional property that there is a number
$\epsilon>0$ such that $f(x)f'(x)>0$  for $a<x<a+\epsilon$.
If there are smooth functions $p$, $q$,  and $r$
defined on $(a,b)$ such that $p(x)r(x)>0$ and
\begin{equation}\lb{zde}
p(x)f''(x)=q(x)f'(x)+r(x)f(x)
\end{equation}
on the interval $(a,b)$, then $f$ is strictly monotone on $[a,b)$.
In particular, $f(x)$ has the same sign on $(a,b)$ that
it does on $(a,a+\epsilon)$.
\end{lem}

The function $f$ defined by~\eqref{fe} satisfies a differential
equation of the form~\eqref{zde}
with $x=e$, $w:=\sqrt{1-e^2}$, and
\begin{align*}
p(e):=&e^4w^2(5w+2),\\
q(e):=&e^3(15w^3 +4w^2+2),\\
r(e):=&e^2(5w^5-8w^4-16w^3+26w^2+24w+4).
\end{align*}
To test the sign of $p(e)r(e)$,
we change the variable to $w$ and note that
$0<w<1$. Let
\[
p^*(w):=p(\sqrt{1-w^2}),\quad r^*(w):=r(\sqrt{1-w^2}).
\]
Clearly, $p^*$ is positive on $0<w<1$. The second factor of
$r^*$ is easily shown to be positive on the same interval.
For example, the second factor is positive at $w=0$ and has no
roots in the interval. The fact that there are no roots can
be checked by computing a Sturm sequence (cf.~\cite{hen}).
This proves $p(e)r(e)>0$ for $e\in I_0$.

To complete the proof of this case, it remains only to
show that there is some $\epsilon>0$ such that
$f(e)<0$ and $f'(e)<0$ on the interval
$0<e<\epsilon$. To do this, we simply note that the Taylor series of
$f$ and $f'$ at $e=0$ are given by
\[
f(e)=-\frac{7}{48}e^5+O(e^7),\quad f'(e)=-\frac{35}{48}e^4+O(e^6).
\]

The fact that $A_1+B_1$ is negative on the interval $I_0$ can
be proved in a similar manner.

We must show that $\partial A_1/\partial G$ has exactly one simple
root on the interval $I_0$.
To do this it suffices to
prove that the function given by
\[
e\mapsto (6-e^2)J_2(e)+e(2e^2-3)J_1(e)
\]
has exactly one simple zero on $I_0$.
Equivalently, using the fact that $J_2(e)$ does not vanish on
$I_0$, it suffices to show that the function
\[f_0(e):=\frac{6-e^2}{2e^2-3}+e\frac{J_1(e)}{J_2(e)}\]
has the same property.
This fact follows from the expression~\eqref{parAB} for
$\partial A_1/\partial G$,
the recurrence formula
\begin{equation}\lb{newrec}
\nu J_\nu(x)-xJ_\nu'(x)=xJ_{\nu+1}(x),
\end{equation}
that is a simple consequence of equation~\eqref{A2} of
Appendix A,
and the connection between $e$ and
$G$.

The fact that $f_0$ has exactly one simple zero is
a consequence of the following three propositions:
$(a)$ The function $f_1(e):=(6-e^2)/(2e^2-3)$ is monotone decreasing on
$I_0$.
$(b)$ The function $f_2(e):=eJ_1(e)/J_2(e)$ is monotone decreasing on
$I_0$.
$(c)$ The function $f_0$ has a zero in $I_0$.

Statement $(a)$ is immediate: $f_1'$ is negative on the interval.
Statement $(b)$ follows from the product representation of the
Bessel function $J_\nu$ given by~\cite{as}
\[
J_\nu(x)=\frac{x^\nu}{2^\nu\Gamma(\nu+1)}\prod_{s=1}^\infty
\big(1-\frac{x^2}{j_{\nu,s}^2}\big),
\]
where $j_{\nu,s}$ denotes the $s$th zero of $J_\nu$ and the fact that
the zeros are interlaced as follows:
\[j_{\nu,1}<j_{\nu+1,1}<j_{\nu,2}<j_{\nu+1,2}<j_{\nu,3}<\cdots . \]
Statement $(c)$ follows from two facts:
$\lim_{e\to 0^+}f_0(e)=2$ and $f_0(1)<0$.
The first fact is immediate from the above  product representation or
from the Taylor series for the Bessel
functions at $e=0$. To obtain the inequality, we use the
product representation of the Bessel functions to deduce
\[
f_0(1)=-5+4 \prod_{s=1}^\infty\big(1-\frac{1}{j_{1,s}^2}\big)/
\prod_{s=1}^\infty\big(1-\frac{1}{j_{2,s}^2}\big).
\]
Since $j_{2,s}>j_{1,s}>1$ for all $s$, the
quotient of the two products is less than unity, as required.

The fact that  $\partial B_1/\partial G$ has a unique simple
zero is proved using a similar analysis.
Just as before, it suffices to show that the following function
has exactly one simple zero on $I_0$:
\[
\tf(e):=-e^3\frac{J_1(e)}{J_2(e)}+e^4-3 e^2+3.
\]
Both terms are monotone decreasing on $I_0$ with $\tf(0)=3$ and
$\tf(1)<0$.

We claim the zeros of $\partial A_1/\partial G$
and $\partial B_1/\partial G$ do not occur at the same point.
After division by nonzero factors and the substitution $m=1$
in~\eqref{parAB}, it suffices to show that the following functions
do not have a common zero on $I_0$:
\begin{align*}
\wa(e):=&(2(1-e^2)^2+4)J_1(e)-e(6-e^2)J_1'(e),\\
\wb(e):=&-3(1-e^2)J_1(e)+e((2-e^2)(1-e^2)+1)J_1'(e).
\end{align*}
If the functions do have a common zero, then the function
\[
\wf(e):=\wa(e)-e\wb(e)
\]
has at least one zero on $I_0$. To prove the claim, we
show that $\wf$ is negative on $I_0$.

Using the recurrence formulas for Bessel's functions, we find that
\[
\wf(e)=(-e^6+3e^4+e^3-3e^2-6e)J_0(e)+(e^5+2e^4-6e^3-5e^2+6e+12)J_1(e).
\]
To prove this function is negative on $I_0$ we will apply
Lemma~\ref{cjlem}. We find that $\wf$ satisfies a differential
equation of the specified form with
\begin{align*}
\wp(e):=& e^2(e^{10}-7e^8-4e^7+31e^6+14e^5-55e^4-45e^3+84e^2+9e-36),\\
\wq(e):=& e(11e^{10}-63e^8-32e^7+217e^6+84e^5-275e^4 -180e^3+252e^2+18e-36),\\
\wr(e):=&-e^{12}-27e^{10}+12e^9+95e^8+11e^7-235e^6\\
             &+273e^5-19e^4-468e^3+351e^2+90e-108.
\end{align*}
Using Sturm sequences, it can be proved that
$\wp(e)\wr(e)>0$ for $e\in I_0$. Moreover, we find that
\[
\wf(e)=-\frac{3}{4}e^3+O(e^4),\quad\wf(e)\wf'(e)=\frac{27}{16}e^5+O(e^6).
\]
This completes the proof of the claim.

In case $m>1$, it suffices to consider the range of the function
$F_m$ given by
$e\mapsto (\partial A_m/\partial e)/(\partial B_m/\partial e)$.
A computation shows that the Taylor series of both the numerator and
the denominator of $F_m$ is given by
$-5 e+O(e^2)$ in case $m=2$ and, in case $m>2$, by
\[
\frac{8m^m(m-1)(m-2)}{2^mm!m^2}e^{m-3}+O(e^{m-2}).
\]
It follows that $\lim_{e\to 0^+}F_m(e)=1$.

We claim that
$e\mapsto \partial B_m/\partial e$ has at least one zero on $I_0$.
If not, then $B_m$ is a monotone function of $e$. A computation
shows that $B_m$ has a removable singularity at $e=0$ and that
\[
B_m(e)=\frac{8m^m(m-1)}{2^mm!m^2}e^{m-2}+O(e^{m-1}).
\]
If $m>2$, then $B_m$ is increasing for $0<e<<1$. But,
from the definition of $B_m$, we have
$\lim_{e\to 1^-}B_m(e)=0$, in contradiction.
For $m=2$, we find that
\[B_2(e)=1-\frac{5}{2}e^2+O(e^4) \]
and $B_2$ decreases for  $0<e<<1$. But, $B_2$ is negative
for $0<<e<1$. To see this just note that near $e=1$, the
sign of $B_2$ is determined by $-J_2'(2e)$. By standard properties
of the Bessel functions (cf.~\cite{as}), $J_2(x)$  is positive
on the interval $(0,j_{2,1}')$, where $j_{\nu,s}'$
denotes the $s$th zero of $J_\nu'$.
Since $\nu\le j_{\nu,s}'$, we have that $-J_2'(2e)<0$ for $0<< e<1$.
Again, since $\lim_{e\to 1^-}B_2(e)=0$, we have
a contradiction.
This proves the claim.

Suppose for the moment that $\partial B_m/\partial e>\partial A_m/\partial e$
on the interval $0<e<1$ and consider the first zero $e_*$ of
$e\mapsto \partial B_m/\partial e$. It follows that
$\partial A_m/\partial e(e_*)<0$
while  $\partial B_m/\partial e>0$ on the interval $0<e<e_*$. Thus,
we have
$\lim_{e\to e_*^-}F_m(e)=-\infty$ and the range
of $F_m$ contains the interval
$(-\infty,1]$, as required.

To complete the  proof, it suffices to show that the function $G_m$
given by
\[
e\mapsto m^2 e^3\big( \partial B_m/\partial e-\partial A_m/\partial e\big)
\]
for $m>1$ is positive on the interval $0<e<1$.
This fact follows from Lemma~\ref{cjlem}.
The Taylor series of $G_m$ at $e=0$ is given by
\[
G_m(e)=\frac{(5m+2)m^m}{2^mm!(m+1)}e^{m+4}+O(e^{m+5}).
\]
Thus, it follows that $G_m(e)G_m'(e)>0$ for $0<e<<1$. We also find that
there are functions $p_m(e)$, $q_m(e)$ and $r_m(e)$
such that $p_m(e)r_m(e)>0$ and
\[p_m(e)G_m''(e)=q_m(e) G_m'(e)+r_m(e) G_m(e)\]
on the interval $0<e<1$.
In fact,
\begin{align*}
p_m:=&
e^3{w}^{6}
\big(5 {w} ^{5}{m}^{3}+(-8w^4+30 {w}^{2}){m}^{2}
+(4 {w}^{3 }+24w)m+8\big), \\
q_m:=&
e^2{w}^{4}\big ((25 {w}^{7}-10 {w}^{5 } ){m}^{3}
+ (-32 {w}^{6}+68 {w}^{4}+30 {w}^{2}){m}^{2}\\
&+ (12 {w}^{5}+24 {w}^{3}+48 w )m+24\big ),\\
r_m:=&
e{w}^{3}m\big (5 {w}^{10}{m}^{4}
 +(-18 {w}^{9}+60 {w}^{7}){m}^{3}+ (-12 {w}^{8}
-52 {w}^{6} +180 {w}^{4} ){m}^{2}\\
&+ (30 {w}^{7}-76 {w}^{5}
+166 {w}^{3})m-12 {w}^{6}+32 {w}^{4}-12 {w}^{2}+24\big ).
\end{align*}
(To verify that $G_m$ satisfies the second
order differential equation with these coefficients, we compute
the derivatives of $G_m$ and then convert all the expressions to
the variable $w$.) Finally, to show $p_m(e)r_m(e)>0$, it suffices to
show that the inequality holds for $0<w<1$. To do this,
view $p_m$ and $r_m$ as
polynomials in $m$ and note that all their
coefficients are positive functions
of $w$.
\end{pf}

\subsection{The $(1:1)$ Resonance}
The fundamental physical result of this section
is the following proposition:
Among the periodic Keplerian orbits in $(1:1)$
resonance with an incident
gravitational wave, there are generally a (nonzero) finite number of
continuable periodic motions. In fact,
the frequency of the gravitational wave fixes the
semimajor axis, $a$,  while the amplitudes and the
phase shift, $\alpha$, $\beta$, $\rho$, of the wave
fix the eccentricity
of the unperturbed Keplerian orbits that are excited
by the perturbation. The inclination of the major axis and
the angular position on the ellipse that complete the set
of initial conditions for a continuable orbit on the excited
ellipse are given by formulas presented below.
However, two facts complicate the
mathematical analysis:  there are  exceptional choices of the wave
amplitudes $\alpha$ and $\beta$ such that none of the periodic
orbits in  $(1:1)$ resonance with the incident
gravitational wave are continuable and there are zeros of the
bifurcation function that are not simple.

The precise mathematical result that we will prove
requires a genericity
assumption. For this we will say that a property of the zero
set of~\eqref{B} is generic relative to the parameter vector
$(\alpha,\beta,\rho)\in \bbR^3$, if it holds for an open and dense subset
of $\bbR^3$.
\begin{prop}\lb{11res}
If $m=1$, then, generically relative to the parameters
$(\alpha,\beta,\rho)$, the zero set
of system~\eqref{B}
is a nonempty finite set consisting entirely of simple zeros.
If $m=1$, $\alpha^2+\beta^2\ne 0$, and $\alpha\beta\sin\rho=0$,
then system~\eqref{B}
has a nonzero finite number of zeros which are all simple.
\end{prop}
\begin{pf}
The first generic assumption is $\calE\calF\ne 0$, the second
generic assumption is that $\alpha^2+\beta^2\ne 0$. (Of course,
if $\alpha^2+\beta^2=0$, then there is no perturbation of the Keplerian
orbits.)
According to Proposition~\ref{zeroprop}, we have $A_1^2-B_1^2\ne 0$ for
$0<e<1$. Thus, if system~\eqref{B} has a solution, then, according
to Proposition~\ref{gbaseprop}, we must have
\[
\big(\frac{\partial A_1}{\partial G}+\frac{\partial B_1}{\partial G}\big)^2
\calF^2-
\big(\frac{\partial A_1}{\partial G}-\frac{\partial B_1}{\partial G}\big)^2
\calE^2=0.
\]
Define
\[\kappa:=\frac{\calE-\calF}{\calE+\calF}\]
and note that, after a simple algebraic manipulation and after
taking into account the
obvious fact that the partial derivatives with respect to $G$
can be replaced with no
loss of generality by partial derivatives with respect to the eccentricity $e$,
the last condition is equivalent to the requirement that
\[
\big(\frac{\partial A_1}{\partial e}-\kappa
\frac{\partial B_1}{\partial e}\big)
\big(\frac{\partial B_1}{\partial e}-\kappa
\frac{\partial A_1}{\partial e}\big) =0.
\]
Also, taking into account the fact that $\calE\ge 0$ and $\calF\ge 0$,
our assumption that
$\alpha^2+\beta^2\ne 0$ implies $0<|\kappa|\le 1$.

To find the solutions of system~\eqref{B}, suppose for the moment
that  the equation
\begin{equation}\lb{eqe1}
\frac{\partial A_1}{\partial e}-\kappa
\frac{\partial B_1}{\partial e} =0
\end{equation}
has a solution. For this value of $e$  the third equation of system~\eqref{B}
vanishes provided  $\sin(2g+\ell+\tau)$ and $\sin(2g-\ell+\sigma)$ are both
equal to one or both equal to minus one. In either case,
$\cos(2g+\ell+\tau)$ and $\cos(2g-\ell+\sigma)$ both vanish. Thus, for
all integers $\calM$ and $\calN$ such that
\[
2g+\ell+\tau=\frac{\pi}{2}+2\pi\calM,\quad 2g-\ell+\sigma=
\frac{\pi}{2}+2\pi \calN,
\]
or such that
\[
2g+\ell+\tau=-\frac{\pi}{2}+2\pi \calM,\quad
 2g-\ell+\sigma=-\frac{\pi}{2}+2\pi \calN,
\]
the fixed value of $e$ together with the (nonzero)
finite number of simultaneous solutions
of these last equations with the property that $0\le g,\ell<2\pi$
give a set of solutions of system~\eqref{B}. A similar result
is valid in case
\begin{equation}\lb{eqe2}
\frac{\partial B_1}{\partial e}-\kappa
\frac{\partial A_1}{\partial e} =0.
\end{equation}

To determine the simplicity of these solutions, we must compute the
Jacobian of system~\eqref{B} with respect to the variables $(G,\ell,g)$
at the given solution.
This Jacobian is easily computed by expanding along
the first column of the Jacobian matrix. Up to a nonzero constant
multiple, we find the value of the Jacobian to be
\[
(A_1^2-B_1^2)\Big(\big(\frac{\partial^2 A_1}{\partial G^2}
   +\frac{\partial^2 B_1}{\partial G^2}\big)
\calF\sin(2g+\ell+\tau)+
\big(\frac{\partial^2 A_1}{\partial G^2}
-\frac{\partial^2 B_1}{\partial G^2}\big)
\calE\sin(2g-\ell+\sigma)\Big).
\]
In particular, using the fact that
$e$ is a monotone function of $G$,
it follows that the solution $(G,\ell,g)$ of system~\eqref{B}
is simple provided the corresponding solution $e$
of equation~\eqref{eqe1}, respectively~\eqref{eqe2}, is simple.

To finish the proof, we must determine the existence and simplicity
of the solutions of equations~\eqref{eqe1} and~\eqref{eqe2}.

If either $\alpha=0$, $\beta=0$, or $\sin\rho=0$,
then $\kappa=0$ and  both
equations~\eqref{eqe1} and~\eqref{eqe2} have  unique simple solutions
by Proposition~\ref{zeroprop}. This proves the second statement
of the proposition.

Since the left-hand sides of equations~\eqref{eqe1} and~\eqref{eqe2} are
both analytic functions of $e$, there are at most a finite number
of solutions on the interval $0<e<1$. Moreover, since the map
$(\alpha,\beta,\rho)\mapsto \kappa(\alpha,\beta,\rho)$
is regular on an open and dense subset of $\bbR^3$, if some of the
solutions of one of the equations  are not simple, then there is
an arbitrarily small perturbation of the triplet $(\alpha,\beta,\rho)$
such that the perturbed equations have a finite number of simple zeros.

The existence part of the first statement of the proposition
is a consequence of the following facts. If $\kappa\ne 1$ (a generic
assumption), then
equation~\eqref{eqe2} has at least one zero. To see this, we note
that the function
$e\mapsto {\partial B_1}/{\partial e}$ has value $-3$ at $e=0$ and
has limit $\infty$ as $e\to 1^-$ while the function
$e\mapsto {\partial A_1}/{\partial e}$ has value $-3$ at $e=0$ and
has a finite value at $e=1$. As long as  $\kappa\ne 1$, then
the function
\[
e\mapsto \frac{\partial B_1}{\partial e}-\kappa
\frac{\partial A_1}{\partial e}
\]
has a negative value at $e=0$ and has limit $\infty$ as $e\to 1^-$.
\end{pf}

The proposition does not give a complete description of the
zero set of system~\eqref{B}. However, since this
description is reduced to an investigation of equations~\eqref{eqe1}
and~\eqref{eqe2} that are algebraic combinations of
Bessel's functions, numerical approximations suggest
the following scenario.
If $\kappa\ne 1$, then the function
\begin{equation}\lb{fe1}
e\mapsto \frac{\partial B_1}{\partial e}-\kappa
\frac{\partial A_1}{\partial e}
\end{equation}
has exactly one simple zero on the interval $0<e<1$.
If $\kappa=1$, then~\eqref{fe1}
vanishes at $e=0$ and increases monotonically
to $\infty$ as $e\to 1^-$.
If $\kappa\le 0$,
then the function
\begin{equation}\lb{fe2}
e\mapsto \frac{\partial A_1}{\partial e}-\kappa
\frac{\partial B_1}{\partial e}
\end{equation}
has exactly one simple zero on the interval  $0<e<1$.
There is a number $\kappa_*\approx 0.036$ such that
if $0<\kappa<\kappa_*$, then~\eqref{fe2}
has exactly two simple zeros, while if  $\kappa>\kappa_*$, then~\eqref{fe2}
has no zeros. If $\kappa=\kappa_*$, then~\eqref{fe2} has exactly
one zero which is not simple.
\begin{remark}
In case $\beta=0$, that is the wave is plane polarized in a very
special direction,
it appears that
the zeros of~\eqref{gbase} are all near $e=1$. The smallest
occurs for the $(m:1)=(1:1)$ resonance in case
$\partial A_1/\partial G=0$. Its root is larger than
$e=0.68$. The root is larger for the higher order resonances. This
suggests that only some ``comets'' could remain periodic after perturbation
by a gravitational wave with this particular polarization.
\end{remark}
\subsection{The $(m:1)$ Resonance, $m>1$}
For the $(m:1)$ resonance we will prove that there are
perturbed periodic solutions for the generic $\alpha$, $\beta$ and $\rho$.
This is the content of the next proposition.
\begin{prop}
If $m>1$, then generically, relative to the parameters
$\alpha$, $\beta$ and $\rho$, system~\eqref{B}
has at least one simple zero.
\end{prop}
\begin{pf}
It suffices to consider $\alpha$, $\beta$ and $\rho$ such that
$\calE\calF\ne 0$.  Let $g$ and $\ell$
denote a solution of the equations
\[2g+m\ell+\tau=\pi/2,\qquad 2g-m\ell+\sigma=\pi/2,\]
and note that
with this choice of $g$ and $\ell$, system~\eqref{B} has a solution
provided $G$ , equivalently $e$, is chosen such that
\[
\big(\frac{\partial A_m}{\partial G}
  +\frac{\partial B_m}{\partial G}\big)\calF
+\big(\frac{\partial A_m}{\partial G}
  -\frac{\partial B_m}{\partial G}\big)\calE=0.
\]
Equivalently, as in Proposition~\eqref{11res}, there is a solution provided
\begin{equation}\lb{m1eq}
\frac{\partial A_m}{\partial G}-\kappa
  \frac{\partial B_m}{\partial G}=0.
\end{equation}
Since $|\kappa|<1$,  an application of
Proposition~\eqref{zeroprop} shows that~\eqref{m1eq}
has at least one solution.
Moreover, since $A_m^2-B_m^2$ is not the zero function, it has only
a finite number of zeros for $0<e<1$. Also, its zeros do not depend
on the choice of the parameters $\alpha$, $\beta$ and $\rho$.
Thus, if necessary, after a  perturbation of the
parameters we can be sure that our solution of~\eqref{m1eq} is not a zero of
$A_m^2-B_m^2$ and that it is a simple zero of the
left hand side of~\eqref{m1eq}. As in Proposition~\eqref{11res}, it follows
that the corresponding choice of $(G,\ell,g)$ is a simple zero of
system~\eqref{B}.
\end{pf}
\section{\lb{cpw}Circularly Polarized Waves}
In this section we consider the equations of motion~\eqref{beqn} for the
case of a circularly polarized incident wave. This corresponds to
the special case where, in the components of
the tidal matrix $K$ in~\eqref{bbeqn}, we take
$\alpha=\beta$ and $\rho=\pm \pi/2$. The minus sign
corresponds to a right circularly
polarized wave, while the plus sign corresponds to
a left circularly polarized wave. We note that this is the
main case excluded from the analysis of the previous section. There,
the bifurcation function does not have simple zeros for the bifurcation
problem corresponding to circular polarization.

The equations of motion for the right circularly polarized wave have
the form
\begin{align*}
\frac{d^2 x}{dt^2}+\frac{kx}{r^3}=&-\epsilon\alpha\Omega^2
            (x\cos\Omega t+y\sin\Omega t),\\
\frac{d^2 y}{dt^2}+\frac{ky}{r^3}=&\epsilon\alpha\Omega^2
            (-x\sin\Omega t+y\cos\Omega t).
\end{align*}
This system can be treated in a similar manner as the analogous system
that arises in
Hill's lunar theory (cf.~\cite{Kov}~\cite{hp}~\cite{Stern}).
The key idea is to view the system in a new Cartesian coordinate
system that rotates relative to the inertial system
with half the frequency of the gravitational wave.
This factor of $1/2$ is due to the fact that the wave has helicity $+2$.
In these rotating coordinates,
that we again call $x$ and $y$,
we obtain  the following  system
\begin{align}
\notag\frac{d^2 x}{dt^2}-\Omega\frac{dy}{dt}
  -(\frac{1}{4}-\epsilon \alpha) \Omega^2 x +\frac{kx}{r^3}&=0,\\
\lb{rsys}\frac{d^2 y}{dt^2}+\Omega\frac{dx}{dt}
  -(\frac{1}{4}+\epsilon \alpha) \Omega^2 y +\frac{ky}{r^3}&=0.
\end{align}
We note that the replacements $t\to -t$ and $\Omega\to-\Omega$
leave the system invariant. Thus, it suffices to consider the
equations of motion for either state of circular polarization.
\begin{figure}[t]
\caption[]{
Orbits of a Poincar\'e map for the Hamiltonian system with
Hamiltonian~\eqref{hamrp}. The parameters are
$\alpha=1$, $\Omega=1$, $k=1$, $\epsilon=0$ (left hand panel),
$\epsilon=.026$ (right hand panel), and
the energy is $H(p_r,p_\theta,r,\theta)=H(0,1,1,0)$.
For post script figures contact
carmen\@chicone.cs.missouri.edu
\label{fig1}}
\end{figure}

A remarkable fact, also utilized by Hill,
is that system~\eqref{rsys} is equivalent to a
Hamiltonian system with Hamiltonian
\[
H=\frac{1}{2}(X^2+Y^2)+\frac{\Omega}{2}(yX-xY)-\frac{k}{r}+
\frac{\epsilon}{2}\alpha\Omega^2(x^2-y^2),
\]
where $X:=\dot x-\Omega y/2$, and $Y:=\dot y+\Omega x/2$.
This Hamiltonian is given in polar coordinates by
\begin{equation}\lb{hamrp}
H=\frac{1}{2}(p_r^2+\frac{p_\theta^2}{r^2})-\frac{k}{r}
   -\frac{\Omega}{2}p_\theta
   +\frac{\epsilon}{2}\alpha\Omega^2r^2 \cos 2\theta,
\end{equation}
where
$p_r=(xX+yY)/r$ and $p_\theta=xY-yX$,
and in Delaunay elements by
\[
H=-\frac{k^2}{2L^2}-\frac{\Omega}{2}G+\frac{\epsilon}{2}\alpha\Omega^2
\calC(L,G,\ell,g).
\]

The Delaunay form of the Hamiltonian is expressed in action-angle
variables and is in the correct form to apply the
Kolmogorov-Arnold-Moser~(KAM) theory
(see for example~\cite{ArnoldCM}~\cite{ds3}).
Here, the Hamiltonian is degenerate. But, as pointed out in
Sternberg~\cite[Vol. 2, p. 257]{Stern}, the Hamiltonian system
with Hamiltonian $H^2$ has the same trajectories as the original
system, only the speed along trajectories is changed.
Moreover, the unperturbed part of $H^2$ satisfies the nondegeneracy
assumption for the KAM theorem---its Hessian, with
respect to the actions,  has a nonzero determinant. Thus, the
perturbed trajectory remains bounded in time, being
trapped between two-dimensional invariant tori in the three-dimensional
energy surfaces of our {\em two-degree} of freedom
Hamiltonian. Thus, sufficiently weak
circularly polarized gravitational waves do
not ``ionize'' the Keplerian ellipses; that is,  the
osculating semimajor axes do not become unbounded. This is illustrated
in Fig.~\ref{fig1}, where  ``phase portraits'' for a
typical Poincar\'e map for the Hamiltonian
system corresponding to~\eqref{hamrp} is depicted.
After an energy $H_0$ is fixed,  each orbit
on the graph is produced by first choosing an initial
point $(p_r,r)$ in the depicted plane and then by marking the
position of the $(p_r,r)$ coordinates of the Hamiltonian orbit,
with initial condition $(p_r,r)$, $\theta=0$, and with $p_\theta$
the implicit solution of
$H(p_r,p_\theta,r,0)=H_0$, at each time when $\theta(t)$ is
a multiple of $2\pi$ and $\dot\theta(t)\dot\theta(0)>0$. In the actual
computation, $\theta$ is reset to zero each time a crossing is marked.
The figure contrasts the foliation by invariant tori for the
unperturbed system, where there appears to be an
incidental resonance of order two
and one of order three,
with the existence of several invariant tori, as
well as a strongly stochastic layer, for the perturbed Poincar\'e map.

We mention that the existence of periodic solutions of the
equations of motion in
the rotating coordinate system (these  correspond to periodic
or quasiperiodic motions of the original system) can be
proved along the lines of Poincar\'e's periodic solutions
of the first and the second kind for the restricted three-body problem.

The continuation theory for the periodic solutions of
the first kind does not depend on the perturbation terms, only
on the Floquet multipliers of the  ``circular'' periodic orbits
of the Kepler problem in the rotating coordinate system
(see~\cite{MH}~\cite{Stern}). The unperturbed orbits that continue
are given by $p_\theta=C$, $p_r=0$, $r=C^2/k$ for a fixed constant
$C$.

The continuation theory for the elliptical orbits,
periodic orbits of the second kind, can be completed along the
lines attributed to Poincar\'e and subsequent authors as outlined
in~\cite[Vol 2, p. 274]{Stern}. However, these can also be
found using the continuation theory of \S~\ref{ct}. In the following
brief outline of the procedure, we will use
the Delaunay formulation of the equations of motion.

After isoenergetic reduction,
by implicitly solving for the angular momentum $G$
in the perturbed Hamiltonian as a power series
in $\epsilon$,
and reformulation of the reduced system as a system
of  differential
equations with the timelike independent variable $g$, one obtains
a system of two differential equations that are  $\pi$-periodic
in $g$. The continuation theory of \S~\ref{ct} can  be applied to
this reduced system.

Here, the Poincar\'e section is given by the submanifold
defined by $g=0$, and the return map is an iterate of the strobe after
each $g$ interval of length $\pi$.
An $(m:n)=(m:1)$ resonant unperturbed orbit corresponds to
an invariant one dimensional torus in the Poincar\'e section.
All such tori are  normally
nondegenerate due to the fact that the periods of the unperturbed
orbits, in the reduced unperturbed
system, change monotonically with $L$ (this is equivalent to the
twist condition for the Poincar\'e map).
The corresponding bifurcation function maps
the angular variable $\ell$ along the unperturbed orbit
to the average of the first order part of the reduced
differential equation for the action $L$ over the unperturbed
resonant orbit with initial value $\ell$.
In fact, the function  is given (up to a nonzero constant
multiple) by $\ell\mapsto (A_m(e)-B_m(e))\sin m\ell$
where $A_m$ and $B_m$ are defined in~\eqref{fcoeff}.
This function has simple zeros (for almost all values of
the eccentricity). Hence, the unperturbed resonant
orbits have continuations.
In particular, our method  produces a periodic
orbit of the form $g\mapsto (L(g,\epsilon),\ell(g,\epsilon))$
for the reduced system with independent variable $g$.

Using the fact that $G$ is implicitly given as a function
of the form
\[G:=G(L(g,\epsilon),\ell(g,\epsilon),g,\epsilon),\]
we see that $G$ is also periodic in $g$. Finally, to obtain
$g$ as a function of time, we use the first order differential
equation
\[
\frac{dg}{dt}=-\frac{\Omega}{2}+\epsilon\frac{1}{2}\alpha\Omega^2
\frac{\partial \calC}{\partial G}
(L(g,\epsilon),G(g,\epsilon),\ell(g,\epsilon),g).
\]
This last equation, at least for sufficiently small $\epsilon$,
has solutions $g(t)$ such that, for some period $T(\epsilon)>0$,
its solution satisfies
$g(t+T(\epsilon))=g(t)-2\pi$. Since $g$ is an angular variable,
the corresponding function
\[t\mapsto (L(g(t),\epsilon),G(g(t),\epsilon),\ell(g(t),\epsilon),g(t))\]
produces a periodic solution of the original perturbation problem
in the rotating coordinate system. These solutions are analogous
to Poincar\'e's periodic solutions of the second kind.

\section{\lb{scn}Speculations, Conjectures and Numerics}
Will a Keplerian binary perturbed by an incident gravitational wave ionize?
To make this question precise, we consider the unperturbed system to be
a Keplerian ellipse, that is, the eccentricity $e$ of the unperturbed
orbit satisfies $0\le e<1$; equivalently, the energy  of the
unperturbed system defined
by Hamiltonian~\eqref{KepHam} is negative.
The corresponding perturbed orbit
(the Hamiltonian trajectory given by~\eqref{perHam})
generally does not lie on an ellipse.
However, we define its osculating conic section
at the instant the perturbed motion reaches the state
$(p_r,p_\theta,r,\theta)$
to be the conic obtained as the Keplerian orbit with this initial data, that
is the Keplerian motion that would be obtained if the perturbation were
``turned off'' at this instant of time.
To ionize, the flow of energy between the binary and the wave
must turn unidirectional in a time averaged sense
in the course of the perturbation such that the binary would steadily
absorb energy; in time, the binary system would be permanently disrupted
and the two bodies would eventually be infinitely far apart from each
other.
On the other hand, the basic equation of motion~\eqref{beqn} breaks
down once the semimajor axis of the osculating ellipse becomes
comparable to the (reduced) wavelength of the incident gravitational
wave. To study the route to ionization, we therefore
introduce the notion of dissociation.
We say the Keplerian ellipse determined by the initial data
$(p_r,p_\theta,r,\theta)$ at time $t_0$ {\em dissociates}
under the influence of the perturbation if at some later time the
osculating conic along the perturbed orbit is a hyperbola.
Equivalently, if one wishes to
remove the geometric language of this definition, the requirement for
dissociation may be recast as follows:
the Keplerian energy $H(p_r(t),p_\theta(t),r(t),\theta(t))$, where
$H$ is given by~\eqref{KepHam}, defined along the perturbed orbit
becomes positive in the course of time.

The ionization question probably does not have a simple answer.
However,
two facts are clear. If the strength of the perturbation is sufficiently
small, there are Keplerian binaries that do not ionize. Independent of
the strength of the perturbation, there are Keplerian binaries that do
dissociate. The first fact is proved in this paper: some of the resonant
Keplerian orbits continue to periodic orbits of the perturbed system.
We also recall that in the case of sufficiently weak
circularly polarized incident
gravitational waves, as discussed in~\S~\ref{cpw}, none of the
Keplerian orbits ionize.
On the other hand, to see that dissociation is possible and
to speculate on the fate of
all orbits, we must review the geometry of our problem.

Recall that Hamiltonian~\eqref{perHam} defines a
$2\frac{1}{2}$-degree of freedom Hamiltonian system.
This system is equivalent to the three-degree of freedom
Hamiltonian system given by
\begin{equation}\lb{3deg}
\calH^*(J,p_r,p_\theta,s,r,\theta)=J+
    \frac{1}{2}(p_r^2+\frac{p_\theta^2}{r^2})-\frac{k}{r}
+\epsilon r^2 (\phi(s)\cos{2\theta}+\psi(s)\sin{2\theta}),
\end{equation}
where $J$ is a ``fictitious'' action variable conjugate to the ``time''
$s$.
Note here that the phase space of~\eqref{3deg}
is six dimensional and that the
five dimensional submanifold $\calP$ given by
\[
H(p_r,p_\theta,\theta,r)=\frac{1}{2}(p_r^2+\frac{p_\theta^2}{r^2})-\frac{k}{r}=0
\]
separates the phase space.
This submanifold corresponds to the parabolic
Kepler orbits while a Keplerian binary corresponds to an initial point
in the region of the six dimensional phase space given by
$H(p_r,p_\theta,\theta,r)<0$. Dissociation occurs provided the perturbed
orbit of the Keplerian binary
eventually crosses the manifold $\calP$.

To determine that some
perturbed orbits do in fact cross the manifold $\calP$,
in both directions, we simply compute the derivative of $H$ along the
perturbed orbit to obtain
\[
\dot H=-2\epsilon\big(
rp_r(\phi(s)\cos 2\theta+\psi(s)\sin 2\theta)+
p_\theta(\psi(s)\cos 2\theta -\phi(s)\sin 2\theta)
\big).
\]
This derivative may be interpreted as a measure of the cosine of the
angle between  the
perturbed Hamiltonian velocity field and the normal to the submanifold
$\calP$. It is apparent that there are open sets on $\calP$ where
$\dot H>0$. Thus, there are open subsets (obtained by reversing the
flow on the boundary set)
of the region
$H<0$ such that every point of the subset corresponds to a Keplerian
ellipse that dissociates. However, we emphasize the fact that time, represented
by $s$ in our three-degree of freedom system,
is one of the variables under
consideration when these open sets are determined.
Thus, the initial data
for a Keplerian ellipse that dissociates include an initial time $t=t_0$.
We do not know, from this analysis, how far back in time the reversed orbits
remain in the region where $H<0$.

A more sophisticated analysis might be based on the diffusion
properties of the
orbits of $\calH^*$. The geometric picture that is believed to
hold for the dynamics of a nearly integrable Hamiltonian system
with at least
three degrees of freedom is easy to describe informally: there
is a dense set of invariant tori coexisting with
a dense set of orbits some of which are
dense in their respective energy
surfaces (see~\cite{ArnoldCM}~\cite{ds3}).

For the
Hamiltonian~\eqref{3deg}, it is easy to see that each five dimensional
energy surface intersects $\calP$. In fact, each energy surface
intersects the subsets of  $\calP$ defined by $\dot H>0$ and $\dot H<0$.
Thus, in the situation of the
conjectured dynamics, for a sufficiently small perturbation strength,
some of the orbits not on invariant tori ionize while a large subset
of the orbits on invariant tori do not. Of course, some of the invariant
tori of the perturbed system might intersect $\calP$; under our definition,
the corresponding orbits will dissociate even though these
same orbits will repeatedly
return to the region where their osculating conics are ellipses.

We note that the usual theory that is used to
prove the existence of invariant
tori for nearly integrable Hamiltonian systems, namely the KAM theory,
is not directly applicable to the Hamiltonian given by~\eqref{3deg} because
the unperturbed system does not meet the required nondegeneracy
conditions. In fact, this Hamiltonian is degenerate
and isoenergetically degenerate (cf.~\cite[p. 408]{ArnoldCM}).
In particular, these facts are evident from the Delaunay action-angle
coordinates for~\eqref{3deg}, where
we see that the three-dimensional unperturbed
invariant tori given by fixing
$J$, $L$, and $G$ do not even have dense trajectories because $\dot g=0$.
To obtain
``nondegenerate'' tori, one must consider
the two dimensional tori given by fixing $J$, $L$, $G$,
and  $g$ while leaving $\ell$ and $s$ free. Some of these tori
may survive perturbation and, given their dimensions, it is possible
that some of the perturbed tori are ``whiskered'': they have
stable and unstable manifolds (each with dimension at most three).
The existence of these invariant manifolds---together with
the stable
and unstable manifolds associated with periodic solutions
(one dimensional invariant
tori) and the
intersections among them---is likely responsible
for the diffusion of some of the orbits not in the union of
these invariant sets and the invariant tori.
 \begin{figure}[th]
 \caption[]{
 Projections into $(p_r,r)$ plane of part of one orbit, approximately
5000 iterates in each panel,
 of the Poincar\'e map
 for the Hamiltonian system with
 Hamiltonian~\eqref{perHam}. The parameters are
 $\alpha=\beta=2$, $k=1$, $\rho=0$,
 and $\epsilon$ in the panels from left
to right is $0.0$, $0.001$, $0.002$ and $0.0025$.
 The initial values are
 $(p_r,p_\theta,r,\theta)=(.5, 1, 1, 0)$. In this case,
 $\Omega$ is chosen ($\Omega\approx 3.897$) so that
 the unperturbed Keplerian ellipse has frequency approximately
 $1/6$th the frequency of the incident gravitational wave.
The region bounded by the branches of the curve given by $rp_r^2=2k$ shown
in the panels corresponds to elliptical motion.
To obtain the .ps files for this figure or hard copies of the figure,\\ contact
 carmen\@chicone.cs.missouri.edu
 \label{fig2}}
 \end{figure}

We end this section with a short description of some of the numerical
experiments performed on the Hamiltonian system~\eqref{perHam}
given by
\begin{equation*}
\calH(p_r,p_\theta,r,\theta)=
    \frac{1}{2}(p_r^2+\frac{p_\theta^2}{r^2})-\frac{k}{r}
+\epsilon r^2 (\phi(t)\cos{2\theta}+\psi(t)\sin{2\theta}),
\end{equation*}
where $\phi$ and $\psi$ are given by equation~\eqref{phipsi}.
The results of a typical experiment that suggests the possibility of
dissociation for an elliptical Keplerian orbit
with eccentricity $e=0.5$ are depicted in
Fig.~\ref{fig2}. To obtain the figure,
the above $2\frac{1}{2}$-degree of freedom
Hamiltonian system is integrated numerically, and the values of
the  solution are projected into the  $(p_r,r)$-plane
after each time interval of length $2\pi/\Omega$. The figure
suggests that dissociation will occur for values of $\epsilon$ that
exceed $\epsilon\approx 0.002$.

To gain some insight into the absorption of gravitational waves by
a Newtonian binary, we have also tested the ``rate of dissociation'', defined
to be inversely proportional to
the number of iterates of the Poincar\'e map required before
the osculating conic of the perturbed ellipse becomes a hyperbola,
by numerical integration of the Hamiltonian system for various
values of the frequency $\Omega$ and
phase shift $\rho$ of the incident wave. We assume here that
$\alpha=\beta$; moreover, the initial elliptical motion is
counterclockwise.  Although these experiments are somewhat
difficult to interpret, one fact seems to emerge
regarding the sensitivity of the rate of dissociation to
the polarization of the wave. For fixed $\Omega$,
the maximal dissociation rate is in the vicinity of $\rho=-\pi/2$ while
the minimal dissociation rate is in the vicinity of $\rho=\pi/2$. This rate
also depends on $\Omega$, but in a seemingly unpredictable manner.

\section{Appendix A: $\calC$ and $\calS$ in terms of Delaunay elements}
The purpose of this appendix is to express $\calC=r^2\cos{2\theta}$ and
$\calS=r^2\sin{2\theta}$ in terms of Delaunay elements. It follows from
the relation $\theta=v+g$ that
\begin{align*}
\calC=&r^2\cos(2g+2v)
     =r^2\cos{2v}\cos{2g}-r^2\sin{2v}\sin{2g},\\
\calS=&r^2\sin(2g+2v)
     =r^2\sin{2v}\cos{2g}+r^2\cos{2v}\sin{2g}.
\end{align*}
Moreover,
\[
\cos v =\frac{\cos \wu-e}{1-e\cos \wu},\quad
\sin v=\sqrt{1-e^2}\,\frac{\sin \wu}{1-e\cos \wu};
\]
these relations follow from~\eqref{anomaly},
and the fact that by definition $v\to \wu$ as
$e\to 0$ (cf.~\cite[Vol.~1, p.~100]{Stern}).
Therefore,
\begin{align*}
r^2\cos{2v}=&a^2(\frac{3}{2}e^2-2e\cos \wu+\frac{1}{2}(2-e^2)\cos{2\wu}),\\
r^2\sin{2v}=&a^2\sqrt{1-e^2}\,(\sin{2\wu}-2e\sin \wu).
\end{align*}

There are classical expansions
for $\cos{j\wu}$ and $\sin{j\wu}$ in Fourier series whose coefficients are
expressible in terms of the Bessel function $J_\nu$ of order $\nu$. Here,
this Bessel function
is most conveniently defined by
\[
J_\nu(x):=\frac{1}{2\pi}\int_0^{2\pi}\cos(\nu t-x\sin t)\,dt.
\]
Following, for example, J.~Kovalevsky~\cite[p.~49]{Kov},
one finds that
\begin{align*}
\cos \wu=&-\frac{e}{2}
        +\sum_{\nu=1}^\infty\frac{1}{\nu}
      [J_{\nu-1}(\nu e)-J_{\nu+1}(\nu e)]\cos{\nu\ell},\\
e\sin \wu=&2\sum_{\nu=1}^\infty\frac{1}{\nu}J_\nu(\nu e)\sin{ \nu\ell},
\end{align*}
and, for $j>1$,
\begin{align*}
\cos j\wu=&\sum_{\nu=1}^\infty\frac{j}{\nu}
        [J_{\nu-j}(\nu e)-J_{\nu+j}(\nu e)]\cos{\nu \ell},\\
\sin j\wu=&\sum_{\nu=1}^\infty\frac{j}{\nu}
[J_{\nu-j}(\nu e)+J_{\nu +j}(\nu e)]\sin{\nu \ell}.
\end{align*}

Using these expansions, we obtain the Fourier series
given in~\eqref{CSfs}
where
\begin{align}
\notag A_\nu=&\frac{1}{\nu}
   \big((2-e^2)[J_{\nu-2}(\nu e)-J_{\nu+2}(\nu e)]
               -2e[J_{\nu-1}(\nu e)-J_{\nu+1}(\nu e)]\big),\\
\lb{A1}B_\nu=& \frac{2}{\nu}\sqrt{1-e^2}\,
  \big([J_{\nu-2}(\nu e)+J_{\nu+2}(\nu e)]-2J_{\nu}(\nu e)\big).
\end{align}
We note that the functions
$\calC$ and $\calS$ are analytic and
$2\pi$ periodic in the angle variables  $\ell$ and $g$. Moreover,
partial derivatives with respect to the Delaunay elements
can be obtained by differentiation
of their Fourier series term by term.

To simplify the expressions for the Fourier coefficients computed
above, we will use
the following elementary identities for
the Bessel functions~\cite[p.~48]{Kov}:
\begin{align}
\notag J_\nu(x)=&\frac{x}{2\nu}[J_{\nu-1}(x)+J_{\nu+1}(x)],\\
\lb{A2} J_\nu'(x)=&\frac{1}{2}[J_{\nu-1}(x)-J_{\nu+1}(x)],\\
\lb{A3}J_\nu''(x)=&\frac{1}{4}[J_{\nu-2}(x)-2J_\nu(x)+J_{\nu+2}(x)],
\end{align}
and Bessel's equation
\begin{equation}\lb{A4}
x^2J_\nu''(x)+xJ_\nu'(x)+(x^2-\nu^2)J_\nu(x)=0.
\end{equation}
The final expressions for $A_\nu$ and $B_\nu$ given in~\eqref{fcoeff}
are obtained from~\eqref{A1} using~\eqref{A2}-~\eqref{A4}; in
fact, the formula for $A_\nu$ is obtained by standard methods using~\eqref{A2},
and the formula for $B_\nu$ is derived from the original expressions~\eqref{A1}
and~\eqref{A3} after noticing that $B_\nu$
is proportional to $J_\nu''(\nu e)$
and using Bessel's equation~\eqref{A4}.
\section{Appendix B: A binary influenced by a distant massive third body}
The purpose of this appendix is to explore the possibility
of applying our results to the three body problem. For a binary system
influenced by a distant massive third body, our ``tidal'' approach
results in a limiting case of the celebrated problem of three bodies and
the question is whether the continuation
theory of \S~\ref{ct} would be applicable
in this case. The existence of periodic orbits in the three-body
problem has been established in the classical work of Poincar\'e~\cite{hp}.

We study the effect of a massive body,
metaphorically the  Sun,
on a binary, metaphorically the Earth-Moon system,  where the
Sun is viewed as giving rise to a periodic perturbation of the Earth-Moon
orbit by tidal forces.
To derive the equations of motion that will be considered in this appendix,
we
will model the Earth-Moon-Sun system according to the following scenario.
The motion of the Sun is neglected due to its great mass,
its gravitational attraction brings about the motion of the Earth-Moon
system as a whole on an almost Keplerian orbit about the Sun and its tidal
influence perturbs the orbit of the Moon about the Earth.  It is this latter
motion that constitutes the lunar problem under investigation here.

To obtain the mathematical model, let us consider the equations of motion of
$m_1$ and $m_2$---the Earth and Moon in our approximation, respectively---as
given by~\eqref{BasicEq1}, with a single perturbing body, namely the Sun, with
potential
\begin{equation}\lb{B1}
\Phi(\bX) = - \frac{G_0 M_{\odot}}{|\bX - \bX_{\odot}|}.
\end{equation}
It is useful to transform the equations of motion of $m_1$ and $m_2$ from
${\bX}_1(t)$ and ${\bX}_2(t)$ to the relative coordinates
${\br} = {\bX}_1-{\bX}_2$ and the center-of-mass coordinates
${\bX}_{\cm} = (m_1{\bX}_1 + m_2{\bX}_2)/(m_1+m_2)$.
In terms of these new coordinates, the
solar gravitational attraction involves the relative coordinates ${\br}$ and
${\bR} = {\bX}_{\cm}-\bX_\odot$.
Let us further assume that $\mu = r/R \ll 1$, where $r$ and $R$ denote
the magnitudes of the corresponding vectors, and
consider the expansion of the solar influence in these equations in terms of
$\mu$.  Using the fact that
\[
\frac{{\bR} + \eta {\br}}{|{\bR} + \eta {\br}|^3}
= \frac{{\bR}}{R^3} + \frac{\eta}{R^3}\big({\br}
            -3\frac{{\bR}\cdot{\br}}{R^2} {\bR}\big) + O(\mu^2),
\]
where $\eta$ is a constant parameter with $|\eta|<1$,
the equations of motion reduce to
\begin{align}\label{B2}
\frac{d^2 \bX_{\cm}}{dt^2}  = &
       - G_0 M_{\odot} \frac{{\bR}}{R^3} + O(\mu^2), \\
\label{B3}
\frac{d^2 r^i}{dt^2} + \frac{k r^i}{r^3}  =&  - K_{ij} r^j + O(\mu^2),
\end{align}
where we have introduced the ``tidal matrix" $K_{ij}$ such that
\begin{equation}\label{B4}
K_{ij}  = \frac{G_0 M_{\odot}}{R^3}
           \big(\delta_{ij} - 3\frac{R^iR^j}{R^2}\big).
\end{equation}

We consider the external body (the Sun) to be so massive
($M_{\odot} >> m_1, m_2$) that it is essentially
unaffected by the presence of $m_1$ and $m_2$
(the Earth and Moon, respectively).
Thus, we can take ${\bX}_{\odot} = \bzero$
and therefore the Sun remains fixed at the
origin of the inertial coordinate system
under consideration.
Neglecting terms of order $r^2/R^2$ in~\eqref{B2}, this equation
reduces to the Newtonian two-body equation for the relative
motion of the center-of-mass about the Sun. We take this orbit to be slightly
elliptic (for example, for the Earth-Moon orbit about the Sun, the eccentricity
$e_1$
is approximately $0.017$).
The resulting expression for
${\bR}$ can be substituted
into \eqref{B3} to give the equations describing
the dynamics of the Earth-Moon system in the presence of the Sun.  We further
assume that the relative orbit as well as the center-of-mass motion occurs in
the equatorial plane of the Sun.
This should be a reasonable approximation
as the
Earth-Moon orbital plane makes an angle of approximately $5^\circ$
with the ecliptic
(the ecliptic is essentially the plane of the Earth's orbit around the Sun)
while the
ecliptic makes an angle of approximately $ 7^\circ$ with the
equatorial plane of the
Sun. It is clear that
this ``tidal" approach to the three-body problem is somewhat
different from the  standard ``restricted'' approach; in the latter
case, the mass of the Moon is effectively set equal to zero.

The Earth-Moon orbit about the Sun has a small eccentricity; therefore,
the tidal
matrix in \eqref{B3}  will be determined to first order in the
eccentricity.  To this end, let
$\Omega^2 = G_0 M_{\odot}/a_{\odot}^3$
(with $a_{\odot}$ being the semimajor axis of the Earth-Moon orbit around the
Sun)
and note that the eccentric anomaly is
$\wu \approx \Omega t + e_1 \sin{\Omega t}$,
the true anomaly is
$v \approx \Omega t + 2 e_1 \sin{\Omega t}$, and
$R \approx a_{\odot}(1 - e_1\cos{\Omega t})$.
Using $R^1=R\cos v$ and $R^2=R\sin v$, the
Cartesian components of the tidal matrix are given by
\begin{align}\label{B5}
K_{11}  = & - \Omega^2(\frac{1}{2} + \frac{3}{2} \cos{2\Omega t} -
\frac{3}{2} e_1 \cos{\Omega t} (3 - 7 \cos{2\Omega t})), \notag \\
K_{12}  = & - \frac{3}{2}\Omega^2(\sin{2\Omega t}
         + e_1 \sin{\Omega t}(3+7\cos{2\Omega t})), \notag \\
K_{22}  = &- \Omega^2(\frac{1}{2} - \frac{3}{2} \cos{2\Omega t}
        + \frac{3}{2} e_1 \cos{\Omega t}(5 - 7\cos{2\Omega t})), \\
K_{13}  = & K_{23} = 0.\notag
\end{align}
As the tidal matrix is traceless and symmetric, the above equations determine
all of its elements.

Using \eqref{B3} and \eqref{B5} we can write the
associated Hamiltonian for this system.  Because the motion is taken to be
in the equatorial plane, polar coordinates are convenient.  In
these coordinates the Hamiltonian is
\begin{align}
{\cal H}  = &\frac{1}{2} \left(p_r^2 + \frac{p_{\theta}^2}{r^2}\right) -
\frac{k}{r} - \frac{\Omega^2 r^2}{4} \{ 1 + 3e_1 \cos{\Omega t}
\notag \\
 &\quad + 3 \cos{2\theta}[\cos{2\Omega t}
   - e_1\cos{\Omega t}(4-7\cos{2\Omega t})] \notag\\
 &\quad + 3 \sin{2\theta}[\sin{2\Omega t}
      + e_1 \sin{\Omega t}(3 +7\cos{2\Omega t})]\}. \lb{B6}
\end{align}
Upon expressing the Hamiltonian equations in terms of
intrinsic dimensionless quantities, it becomes clear that the
strength of the interaction between the binary and the third body
is $\Omega^2/\omega^2\ll 1$, but the square root of this perturbation
parameter also occurs in the harmonic terms that render
the Hamiltonian~\eqref{B6}
explicitly time-dependent. In particular, the period of the harmonic
terms becomes unbounded as the perturbation parameter goes to zero.
Therefore, the continuation method of \S~\ref{ct} is
not directly applicable in this case;
in fact, the resolution of this problem is
due to Hill (cf.~\cite{Kov}~\cite{hp}). In Hill's approach, the equation
of relative motion~\eqref{B3} is referred to a Cartesian system of
coordinates $\br'$ that rotates with frequency $\Omega$ with respect
to the inertial system. Let $r^i=S_{ij}{r'}^j$, where the nonzero
elements of the orthogonal matrix $S$ are given by
\[
S_{11}=S_{22}=\cos\Omega t, \quad -S_{12}=S_{21}=\sin\Omega t,\quad S_{33}=1;
\]
then, the equations of motion in the new system are $(r'=r)$
\begin{align}
\frac{d^2x'}{dt^2}-2\Omega\frac{dy'}{dt}-\Omega^2 x'+\frac{kx'}{{r'}^3}=&
      -(K_{11}'x'+ K_{12}'y'),\notag\\
\frac{d^2y'}{dt^2}+2\Omega\frac{dx'}{dt}-\Omega^2 y'+\frac{ky'}{{r'}^3}=&
      -(K_{12}'x'+ K_{22}'y'),\lb{B7}
\end{align}
where $K'=S^TKS$, i.e.,
\begin{equation}
\lb{B8}
K_{11}'=-2\Omega^2(1+3 e_1\cos\Omega t),\quad
K_{12}'=-6\Omega^2 e_1\sin\Omega t,\quad
K_{22}'=-\frac{1}{2}K_{11}',
\end{equation}
and $K_{13}'=K_{23}'=0$. The system~\eqref{B7} is
autonomous for $e_1=0$. In this case, periodic solutions exist as
originally demonstrated by Hill and Poincar\'e (cf.~\cite{hp}).
The continuation of such solutions using $e_1$, $0\le e_1\ll 1$, as
the expansion parameter can be  proved, using the
Implicit Function Theorem, as originally conceived by
Poincar\'e (cf.~\cite{MH}~\cite{Kov}~\cite{hp}) for the restricted
three-body problem.
Of course,  the method of \S~\ref{ct} is also applicable by following
the ideas for isoenergetic reduction as discussed in~\S~\ref{cpw}.

\newpage

\noindent Department of Mathematics,
University of Missouri, Columbia, MO 65211

\noindent Department of Physics and Astronomy,
University of Missouri,
Columbia, MO 65211

\noindent Department of Chemical Engineering,
University of Missouri, Columbia, MO 65211

\noindent E-mail addresses:\\
carmen@@chicone.cs.missouri.edu \\physgrav@@cclabs.missouri.edu\\
chendgr@@mizzou1.missouri.edu

\begin{thebibliography}{xxxxxxxxxx}
\bibitem{as} M.~Abramowitz and I.~Stegun,
Handbook of Mathematical Functions,
National Bureau of Standards, Washington DC, 1968.

\bibitem{ArnoldCM} V.~I.~Arnold,
Mathematical Methods of Classical Mechanics, Grad.\ Texts Math.\
{\bf 60}, Springer-Verlag, 1978.


\bibitem{ccc} C.~Chicone,
A Geometric Approach to Regular Perturbation
Theory With An Application to Hydrodynamics,
To appear in Trans.\  AMS.

\bibitem{cj}  C.~Chicone and M.~Jacobs,
Bifurcation of Limit Cycles from Quadratic Isochrones,
J.\  of Diff. Eqs., {\bf 91}, (1991), 268--326.

\bibitem{ds3}
Dynamical Systems III, Ency.\ Math.\ Sci., Vol.~{\bf 3},
V.~I.~Arnold, Editor,
Springer-Verlag, 1988.

\bibitem{hen} P.~Henrici,
Applied and Computational Complex Analysis,
Vol.~{\bf 1}, Wiley, New York, 1974.

\bibitem{ht} R. A.~Hulse and J. H.~Taylor, Discovery of a
Pulsar in a Binary System, Astrophys.~J.~{\bf 195}, (1975), L51--53.

\bibitem{Kov} J.~Kovalevsky,
Introduction to Celestial Mechanics,
Astrophysics and Space Science Library,
Vol.~{\bf 7}, Springer-Verlag, 1967.

\bibitem{mashoon1}  B. Mashhoon,  Tidal Radiation, Astrophys.\ J.,
{\bf 216}, (1977), 591--609.

\bibitem{mashoon2} B. Mashhoon, On Tidal
Resonance, Astrophys.\ J., {\bf 223}, (1978), 285-298.

\bibitem{MH} K.~R.~Meyer and G.~R.~Hall,
Introduction to Hamiltonian Dynamical Systems and the
N-Body Problem,
Applied Mathematical Sciences,
{\bf 90}, Springer-Verlag, 1992.

\bibitem{hp} H.~Poincar\'e, Les M\'ethodes Nouvelles de la
M\'ecanique C\'eleste, Vols. 1--3, Gauthier-Villars, Paris, 1892--99.

\bibitem{Stern} S.~Sternberg,
Celestial Mechanics, Vols.~ 1--2, W.~A.~Benjamin, Inc., New York, 1969.

\bibitem{ht1} J. H.~Taylor, A.~Wolszczan, T.~Damour, and J. M.~Weisberg,
Experimental Constraints on Strong-Field Relativistic Gravity,
Nature {\bf 355}, (1992), 132--136.
\end{thebibliography}
\end{document}